\documentclass[12pt]{article}
\usepackage{amsmath}
\usepackage{stmaryrd}
\usepackage{graphicx}
\usepackage{amssymb}
\usepackage{graphbox}
\usepackage{url}
\usepackage{xcolor}
\usepackage{algorithm2e}

\usepackage[giveninits=true]{biblatex}
\usepackage{array}
\newcolumntype{?}{!{\vrule width 2pt}}
\makeatletter
\newcommand{\thickhline}{%
    \noalign {\ifnum 0=`}\fi \hrule height 2pt
    \futurelet \reserved@a \@xhline
}
\makeatother

\newtheorem{definition}{Definition}
\newtheorem{proposition}{Proposition}
\newtheorem{theorem}{Theorem}

\newtheorem{example}{Example}
\newtheorem{lemma}{Lemma}

\title{Chaotic mechanism description by an elementary mixer for the
template of an attractor }

\author{Martin Rosalie\\
    UMR5096 Laboratoire Génome et Développement des Plantes,\\
    Université de Perpignan Via Domitia, CNRS, France \\
    \\
\url{martin.rosalie@univ-perp.fr}
}

\begin{document}

\maketitle

\begin{abstract}
    A template describes the topological properties of a chaotic attractor. 
    For attractors bounded by genus--1 torus, a linking matrix describes the topology of the template.
    It has been shown that the template depends on the Poincaré section chosen to perform the topological characterisation: four linking matrices describe four templates of the same chaotic attractor.
    This article presents a framework for deriving the elementary mixer of a template to have a unique way to describe chaotic mechanism and dynamics of a chaotic attractor.
    In this framework, chaotic mechanisms are represented by elementary mixers defined by elementary linking matrix.
    Using concatenation between mixers, a classification of chaotic mechanisms is proposed to categorise them by their size.
\end{abstract}


\vspace{2pc}

% \submitto{\NL}

%\maketitle

\section{Introduction}
\label{sec:introduction}

The topological characterisation of three-dimensional dissipative chaotic attractors method has been introduced in 1980's. 
Its
purpose is to use the properties of the periodic orbits to describe the
topological structure of chaotic attractors. Birman and Williams
\cite{birman1983knotted} are the first to apply this technique on the Lorenz
system \cite{lorenz1963deterministic} where the orbits are considered as the
skeleton of the attractor to describe its template. 
The topological characterisation method and its
applications are described by Gilmore, Lefranc, Letellier and co-workers
\cite{mindlin1990classification, tufillaro1990relative, lesceller1994algebraic,
gilmore1998topological, Gilmore2011, Gilmore2007}.
\begin{figure}[htpb]
    \centering
    \includegraphics[width = \textwidth]{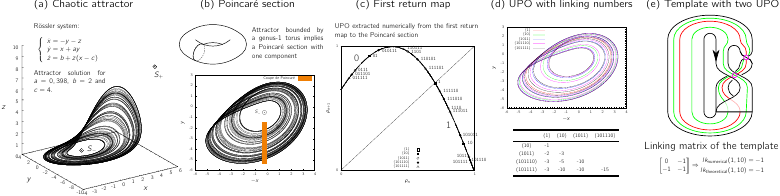}
    \caption{Topological characterisation of a chaotic attractor \cite{letellier1995unstable} (UPO: Unstable Periodic Orbit).
    (a) Three-dimensional representation of the chaotic attractor. (b) The genus--1 bounding torus induces a Poincaré section with one component. (c) First return maps using values from the Poincaré section. (d) Numerical extraction of periodic orbits and their linking numbers. (e) Template of the attractor giving theoretical linking numbers. The template is described with a linking matrix. The template is validated if theoretical and numerical linking numbers are the same. This figure is adapted from Fig.~1 of \cite{rosalie2025}.}
    \label{fig:topo_characterisation}
\end{figure}
To summarise an application of this methodology \cite{letellier1995unstable}, Fig.~\ref{fig:topo_characterisation}
details the topological characterisation of a chaotic
attractor solution to a standard system in the domain: the Rössler system \cite{rossler1976equation}.
The Poincaré section (Figure~\ref{fig:topo_characterisation}b) whose purpose is to transform a continuous problem to a discrete problem with the obtention of the \textit{first return map} that is the signature of the chaotic dynamics (Figure~\ref{fig:topo_characterisation}c).
Unstable Periodic Orbits (UPO) could be considered as the skeleton of a chaotic attractor because of their unstable property (Figure~\ref{fig:topo_characterisation}d).
The Rössler system is numerically solved (using the four-order Runge-Kutta algorithm, RK4) and the solution is a trajectory in a phase space.
Thus, at a moment, this numerical solution will ``follow'' one UPO for a while before diverging and visiting another UPO.
This process keeps the trajectory into the attractor; the latter being the solution of the system that we cannot describe analytically.
The word ``periodic'' in UPO refers to the state space and not to the time space  (topological period).
Periodic orbits are time invariant while the system evolves in a chaotic  state (from initial conditions, the solution evolves and successively visits the unstable
periodic orbits).
% From a chaotic time series (numerical solution of a DES) and with a Poincaré section, UPO can be extracted, and this acquisition is a preliminary step of the \textit{topological characterization} \cite{gilmore1998topological}.
For dissipative systems, the purpose of this method is to obtain the structure of the chaotic mechanism from a topological invariant (the linking number) computed between every couple of UPOs (Fig.~\ref{fig:topo_characterisation}d).
The template is a formal description of the branched manifold detailing the topological properties of a chaotic attractor.
The linking numbers are computed numerically (Fig.~\ref{fig:topo_characterisation}d) and theoretically (from the linking matrix of the template); the template is validated if both are equals for corresponding couple of UPOs (Figure~\ref{fig:topo_characterisation}e).
The non-trivial part of the template is the mixer \cite{rosalie2013systematic} with two strips for this template.
The mixer of the template is described with a linking matrix detailing the number of branches with their torsions and how they permute before joining.
This formalism underlines the chaotic mechanism of the attractor with a splitting chart that separate the flow in branches (sensitivity to initial conditions) and a joining chart where trajectories are regrouped before going for another round (bounded solution).

Algebraic relations are introduced to perform operations between the linking matrices.
Gilmore \textit{et al.} propose the concatenation of linking matrix \cite{mindlin1990classification}, and we recently propose algorithms to perform this concatenation \cite{gilmore2016algorithms}. 
Tufillaro and co-workers \cite{melvin1991template, tufillaro1992experimental} propose another algebraic representation of a linking matrix: only a matrix satisfying the
orientation convention to know how strips are ordered. Using this
representation convention for templates, we present an algebraical relation
between two templates symmetric by inversion \cite{rosalie2013systematic}.
Other representations have been introduced to describe the chaotic mechanism.
For instance, Mart\'in \& Used \cite{martin2009simplified} proposed some links
between templates focusing on how branches are organised after stretching and
folding and before squeezing. Towards the same goal, Cross \& Gilmore
\cite{cross2013dressed} proposed to use return maps to compare chaotic
mechanisms without considering symmetries.

In this article, we first perform the topological characterisation of the Malasoma attractor \cite{malasoma2000simplest} using four distinct Poincaré sections. 
As shown previously \cite{rosalie2015systematic}, this attractor could lead to four templates and there are algebraic relations between the linking matrices of these templates.
These preliminary results are the starting point of this study where conventions are introduced to obtain a template of an attractor from a given Poincaré section.
The next part is dedicated to the algebraical relations between linking matrices when symmetry and torsions are considered. 
Thus, we propose to describe the \textit{chaotic mechanism} by finding a unique linking matrix describing its \emph{elementary mixer} (this is the result of a minimisation problem using an algorithm).
Finally, a classification of chaotic mechanisms is provided for templates composed of at least five strips.

\section{Topological characterisation of the Malasoma attractor}

\begin{figure}[htpb]
    \centering
    \includegraphics[width=0.5\linewidth]{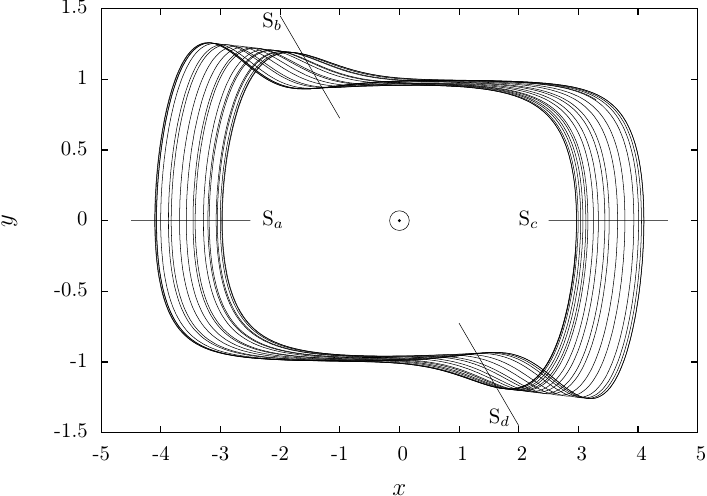}
    \caption{
        Attractor solution to the Malasoma system \cite{malasoma2000simplest} with Poincaré sections \eqref{eq:malasoma_sections}.
    }
    \label{fig:malasoma_01_attra}
\end{figure}

\begin{figure}[htpb]
    \centering
    \begin{tabular}{ccccccc}
    \includegraphics[width=0.2\linewidth]{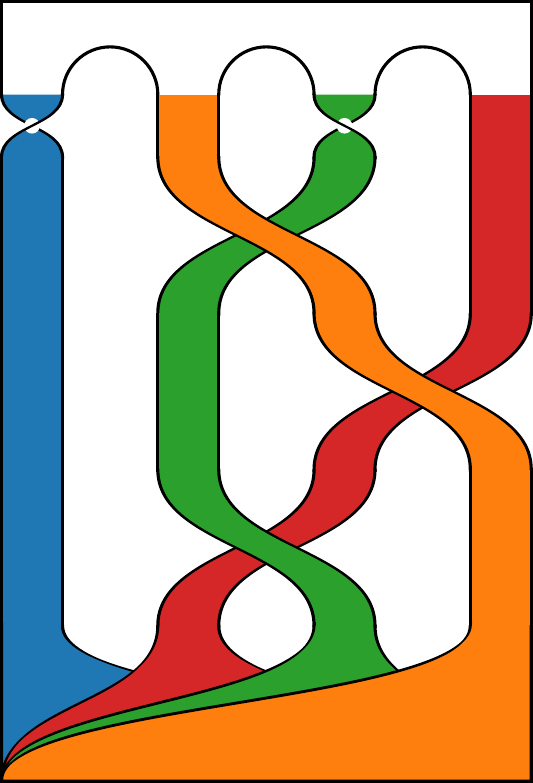} & \quad &
    \includegraphics[width=0.2\linewidth]{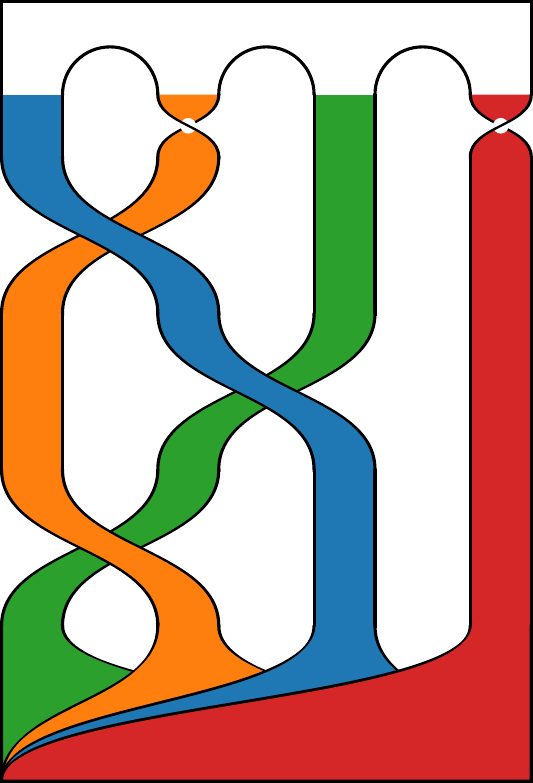} & \quad &
    \includegraphics[width=0.2\linewidth]{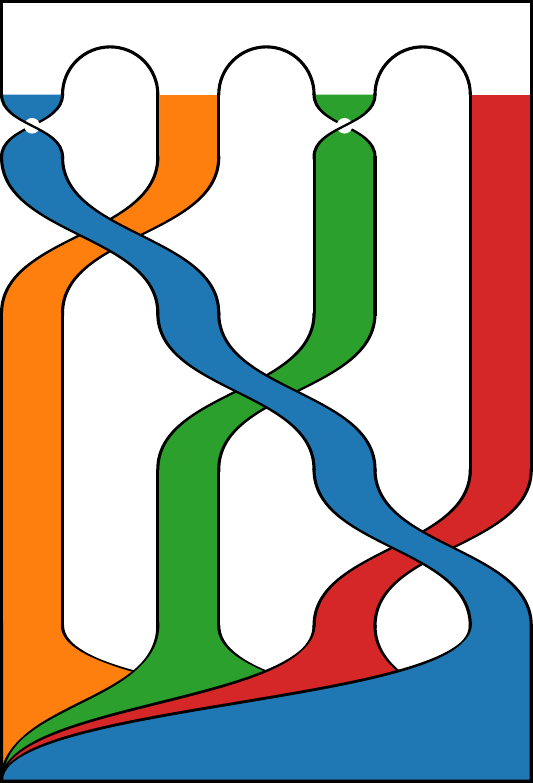} & \quad &
    \includegraphics[width=0.2\linewidth]{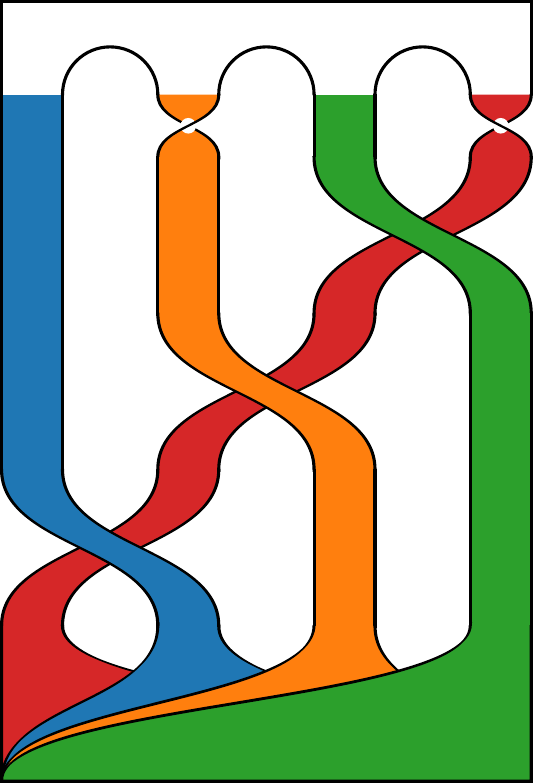} \\
    $L_a$ & & $L_b$ & & $L_c$ & & $L_d$
    \end{tabular}
    \caption{
        Mixers of the template of the attractor depending on the Poincaré
        section (visualisation of the template is done with \texttt{cate} \cite{olszewski2018visualizing}).
    }
    \label{fig:template_abcd}
\end{figure}

The Malasoma system \cite{malasoma2000simplest} has been designed to be the
simplest system producing chaotic dynamics:
\begin{equation}
  \left\{
    \begin{array}{l}
      \displaystyle
      \dot{x} = y \\
      \dot{y} = z \\
      \dot{z} = -\alpha z + x y^2 -x \:,
    \end{array}
  \right.
  \label{eq:malasoma}
\end{equation}
this system is equivariant under an
inversion symmetry. The following section is mainly a synthesis of our
previous results \cite{rosalie2013systematic,rosalie2015systematic} where $(r_n, z_n, \varphi_n)$ are the polar coordinates used to performed the analysis.
Four equivalent Poincaré
sections are defined $\mathrm{S}_i$ ($i \in \{ a, b, c, d \}$) as follow:
\begin{equation}
  \label{eq:malasoma_sections}
  {\cal P}_i \equiv
  \left\{ (r_n,z_n) \in \mathbb{R}^2 ~|~ \theta_n = \varphi_i,
  \dot{\theta}_n < 0
  \right\}  \,
\end{equation}
where $\varphi_a = \pi$, $\varphi_b = \frac{4
\pi}{5}$, $\varphi_c = 0$, and $\varphi_d = \frac{9 \pi}{5}$.
From these Poincaré sections (Fig.~\ref{fig:malasoma_01_attra}),
the topological characterisation method is applied to obtain the template of this attractor. 
This method gives four linking matrix to describe this attractor
\cite{rosalie2015systematic}: $L_x$ where $x=\{a,b,c,d\}$ refers to Poincaré
sections \eqref{eq:malasoma_v4} (Fig.~\ref{fig:template_abcd} is the visualisation of the mixers, non trivial part, of the reduced templates).
\begin{equation}
    \label{eq:malasoma_v4}
\resizebox{.9\textwidth}{!}{ $
    L_a =
    \left[\begin{matrix}
    1 &  0 &  0 &  0\\
    0 &  0 & -1 & -1 \\
    0 & -1 & -1 & -1 \\
    0 & -1 & -1 &  0
    \end{matrix}\right\rrbracket
    \,
    L_b =
    \left[\begin{matrix}
    0 & -1 & -1 & 0 \\
    -1 & -1 & -1 &  0 \\
    -1 & -1 &  0 &  0 \\
    0 &  0 &  0 & 1
    \end{matrix}\right\rrbracket
    \,
    L_c =
    \left[\begin{matrix}
    -1 & -1 & -1 & -1 \\
    -1 &  0 &  0 &  0 \\
    -1 &  0 & 1 &  0 \\
    -1 &  0 &  0 &  0
    \end{matrix}\right\rrbracket
    \,
    L_d = \left[\begin{matrix}
    0 &  0 &  0 & -1 \\
    0 & 1 &  0 & -1 \\
    0 &  0 &  0 & -1 \\
    -1 & -1 & -1 & -1
\end{matrix}\right\rrbracket $
}
\end{equation}

For a better understanding of the topological structure of this attractor, another particular Poincaré sections is considered. The four Poincaré sections \eqref{eq:malasoma_sections} are considered as four components of a unique Poincaré section represented with a unique variable:
\begin{equation}
\rho_n = \left\{
    \begin{array}{ll}
        \frac{r_n - \min(r_n)}{\max(r_n) - \min(r_n)} & \text{ if } (r_n, z_n) \in \mathcal{P}_a \\
        \frac{r_n - \min(r_n)}{\max(r_n) - \min(r_n)} + 1& \text{ if } (r_n, z_n) \in \mathcal{P}_b \\
        \frac{r_n - \min(r_n)}{\max(r_n) - \min(r_n)} + 2& \text{ if } (r_n, z_n) \in \mathcal{P}_c \\
        \frac{r_n - \min(r_n)}{\max(r_n) - \min(r_n)} + 3& \text{ if } (r_n, z_n) \in \mathcal{P}_d \\
    \end{array}
    \right.
\end{equation}
It is usually done for Poincaré section with several components required to analysis attractors bounded by a torus with a higher genus \cite{rosalie2014toward}.
\begin{figure}[htpb]
    \centering
    \includegraphics[width = .5\textwidth]{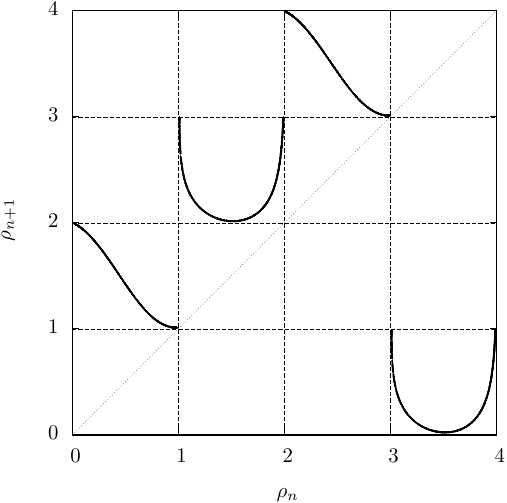}
    \caption{
        First return map to the Poincaré section including four
        components.
    }
    \label{fig:malasoma_2c_appli}
\end{figure}
The first return map (Fig.~\ref{fig:malasoma_2c_appli}) is build from $\rho_n$, it 
details the transitions between the four components. These transitions are either a
decreasing branch or a unimodal shape with a decreasing and an increasing branch.
The decreasing branch shows that in the template the transition is an odd
torsion. The unimodal map indicates that the transition is a mixer with an odd
torsion close the center of the attractor and an even torsion close to the
outside of the attractor. 
A direct template \cite{rosalie2015systematic} details the topological structure of this attractor (Fig.~\ref{fig:template_4_components}).
This attractor is the result of the fusion of two symmetric attractors co-existing for a closer parameter value of $\alpha$ \cite{letellier2014universalities}.

\begin{figure}[tbh]
    \centering
    \includegraphics[width=0.2\linewidth]{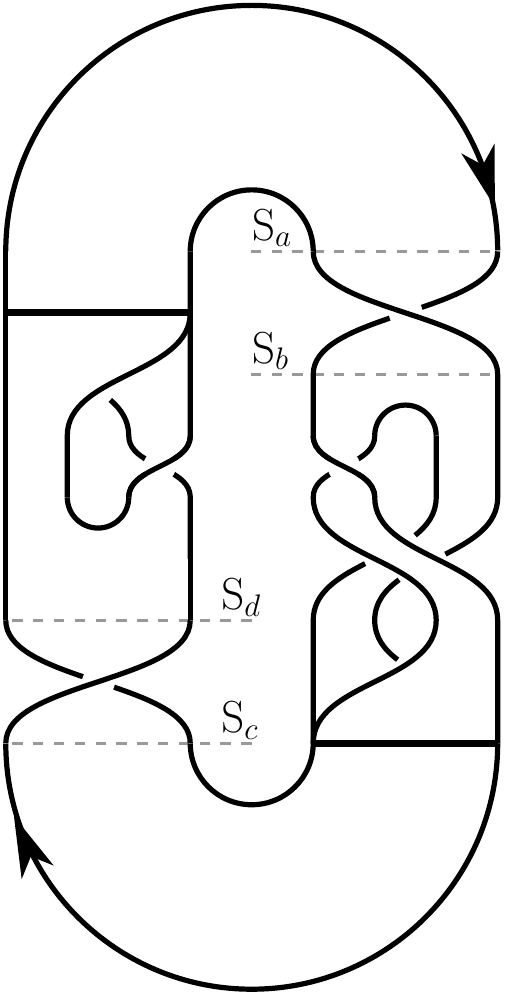}
    \caption{
        Direct template describing the parts of the Malasoma attractor. Figure adapted form Fig.~9 of \cite{rosalie2015systematic}.
    }
    \label{fig:template_4_components}
\end{figure}

\section{Template of chaotic attractors}
\label{sec:templates_of_chaotic_attractor}

\begin{figure}[ht]
  \centering
  \begin{tabular}{ccc}
    \includegraphics[height=8em]{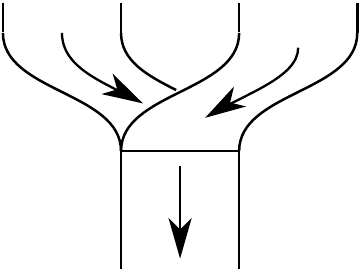} & \quad &
    \includegraphics[height=8em]{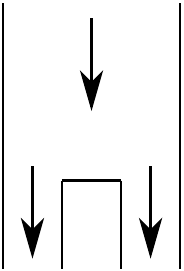}
    \put(-260,38){Branch line $\longrightarrow$}
    \\
    (a) & & (b) \\
  \end{tabular}
  \caption{
    A template is a branched two-manifold with two types of
    charts: (a) joining chart; (b) splitting chart.
  }
  \label{fig:charts}
\end{figure}

\begin{definition} \cite{ghrist1997knots}
  A \emph{template} is a compact branched two-manifold with boundary and
  smooth expansive semiflow built locally from two types of charts:
  joining and splitting. Each chart, as illustrated in Figure
  \ref{fig:charts}, carries a semiflow, endowing the template with an
  expanding semiflow, and gluing maps between charts must respect the
  semiflow and act linearly on the edges.
\end{definition}

\begin{figure}[hbtp]
  \centering
  \begin{tabular}{cccccc}
    \multicolumn{2}{c}{Convention} &
    \multicolumn{2}{c}{Permutations} &
    \multicolumn{2}{c}{Torsions} \\
    \includegraphics[height=4em]{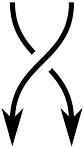} &
    \includegraphics[height=4em]{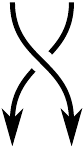} &
    \includegraphics[height=4em]{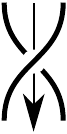} &
    \includegraphics[height=4em]{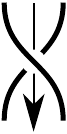} &
    \includegraphics[height=4em]{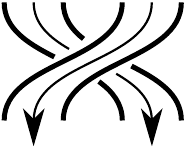} &
    \includegraphics[height=4em]{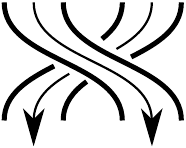} \\
    $+1$ &
    $-1$ &
    positive &
    negative &
    positive &
    negative
  \end{tabular}
  \vspace{-.5em}
  \caption{
    Convention representation of oriented crossings. The permutation
    between two branches is positive if the crossing generated is equal to
    $+1$, otherwise it is a negative permutation. We use the same
    convention for torsions.
  }
  \label{fig:convention}
\end{figure}

A \emph{linker} is a synthesis of the relative organisation of $n$ strips:
torsions and permutations in a planar projection (Fig.~\ref{fig:convention}).
A \emph{mixer} is a linker ended by a joining chart that stretches and
squeezes strips to a branch line.

The orbits are the skeleton of the attractor and structure the flow and
so, the template. The linking number is an integer counting how
many times orbits are entwined one around the other. In the literature, this
method permits to successfully describe a wide range of attractors.  Here we
give some references where the authors give the templates of attractors
bounded by genus--1 torus: R\"ossler attractors by Letellier \textit{et al.}
\cite{letellier1995unstable}, Duffing oscillator attractor by Gilmore \&
McCallum \cite{gilmore1995structure}, Burke-Shaw attractor by Letellier
\textit{et al.} \cite{letellier1996evolution} and recently we propose the
templates of attractors solution to the Malasoma system \cite{rosalie2013systematic,
rosalie2015systematic}. 

In order to compare all templates previously
cited, we already introduce a method that ensures us to obtain a template described by a linking matrix with respect of conventions \cite{rosalie2013systematic}. First, the Poincaré
section used to perform the topological characterisation must be defined.
Then, the template associated to this Poincaré section is described with a
linking matrix with respect to the following conventions:
\begin{enumerate}
    \item Clockwise evolution of the flow.
    \item Poincaré section oriented from the inside to the outside.
    \item Tufillaro's convention \cite{melvin1991template} for joining chart:
        when the strips stretch and squeeze, their order from the left to the
        right corresponds to the bottom to top order.
\end{enumerate}
This definition of the clockwise flow refers to the definition (3.1) of
\cite{franks1985entropy} where only one point is considered around which the
flow evolves clockwise. 
A mixer $\mathcal{M}$
is defined by a linking matrix $M$ where the right side is
``$\;\rrbracket\;$'' to represent the merging at the branch line with respect
to the Tufillaro convention \cite{melvin1991template} (see
\cite{gilmore2016algorithms} to distinguish the two linking matrix
representation conventions).  

\section{Algebraic relation between linking matrices}

For the following section, mixers and linkers (torsions and permutations) of a directed template are considered as \emph{actions}.

\begin{definition}
    In a template $\mathcal{T}$, given two actions $\mathcal{D}_1$ and $\mathcal{D}_2$, when
    $\mathcal{D}_1$ precedes $\mathcal{D}_2$, without any other
    action between, the resulting action is the
    \emph{concatenation} of $\mathcal{D}_1$ before $\mathcal{D}_2$, noted
    $\mathcal{D}_1 \oplus \mathcal{D}_2$.
\end{definition}

\subsection{Concatenation of torsion and mixer}
\label{sub:concatenation_of_torsion_and_mixer}

\begin{lemma}\cite{rosalie2013systematic}
  \label{lema:concatenationTorsionMixer}
  Given a template $\mathcal{T}$, containing a torsion $t$
  defined by $T=|\tau|$ and a mixer $\mathcal{M}$ defined by $M$ with
  $n$ strips. If $\mathcal{M}$ and $t$ are concatenated, then
  \begin{itemize}
    \item $\mathcal{M} \oplus t \equiv \mathcal{M}'$
    \item $t \oplus \mathcal{M} \equiv \mathcal{M}'$, if $\tau$ is
      even
    \item $t \oplus \mathcal{M} \equiv \mathcal{M}''$, if $\tau$ is
     odd
  \end{itemize}
  $\mathcal{M}'$ is defined by $M'$ and $\mathcal{M}''$ is defined by
  $M''$
  \begin{equation}
    M'=\left[
      \begin{matrix}
    M_{1,1}+\tau & \cdots & M_{1,n}+\tau \\
          \vdots & \ddots & \vdots \\
    M_{n,1}+\tau & \cdots & M_{n,n}+\tau \\
      \end{matrix}
      \right\rrbracket\quad
    M''=\left[
      \begin{matrix}
    M_{n,n}+\tau & \cdots & M_{n,1}+\tau \\
          \vdots & \ddots & \vdots \\
    M_{1,n}+\tau & \cdots & M_{1,1}+\tau \\
      \end{matrix}
      \right\rrbracket\,.
  \end{equation}
\end{lemma}

\subsection{Concatenation of mixers}
\label{sub:concatenation_of_mixers}

\begin{theorem}\cite{rosalie2015systematic,gilmore2016algorithms}
  \label{theo:concatMixer}
  Given a template $\mathcal{T}$ containing a mixer $\mathcal{A}$ of $n_a$
  strips defined by the linking matrix $A$ and a mixer $\mathcal{B}$ of $n_b$
  strips defined by the linking matrix B. If $\mathcal{A}$ is concatenated
  before $\mathcal{B}$, then $\mathcal{A} \oplus \mathcal{B} \equiv
  \mathcal{C}$, where $\mathcal{C}$ is a mixer of $n_a\times n_b$ strips defined
  by the linking matrix $C$ that is the sum of the expanded linking matrices
  $A$ and $B$  with additional permutations due to the insertion of
  $\mathcal{A}$
  \begin{equation}
    C = A_\text{expand}
    + A_\text{insertion}
    + B_\text{expand}
  \end{equation}
  with the respect of the strips order at the beginning of $\mathcal{A}$.
\end{theorem}

\begin{figure}[ht]
  \centering
  \begin{tabular}{ccccccc}
    \includegraphics[width=0.168\textwidth]{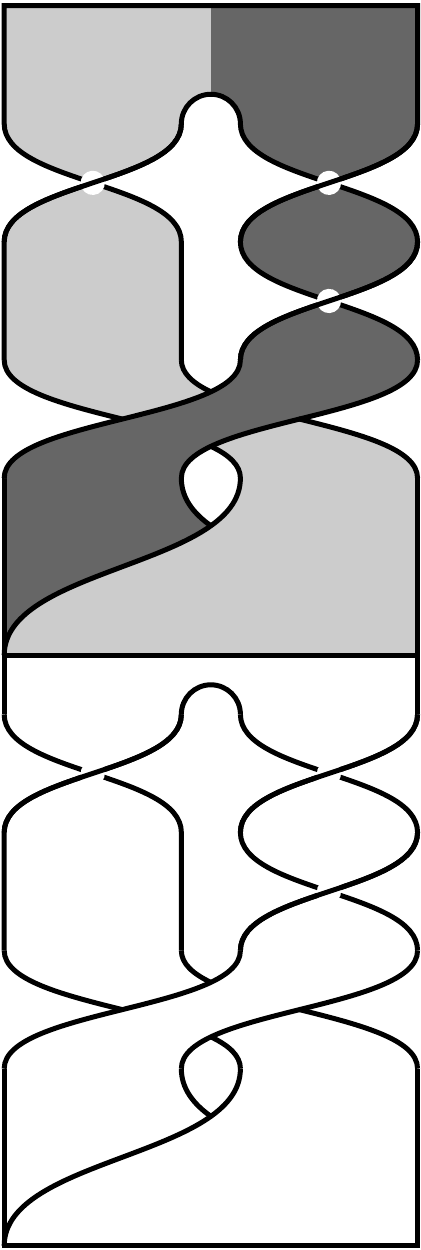}
    \put(-65,215){$s_1$}
    \put(-22,215){\color{white}$s_2$}
    \put(-65,95){$t_1$}
    \put(-22,95){$t_2$}
    & \quad &
    \includegraphics[width=0.168\textwidth]{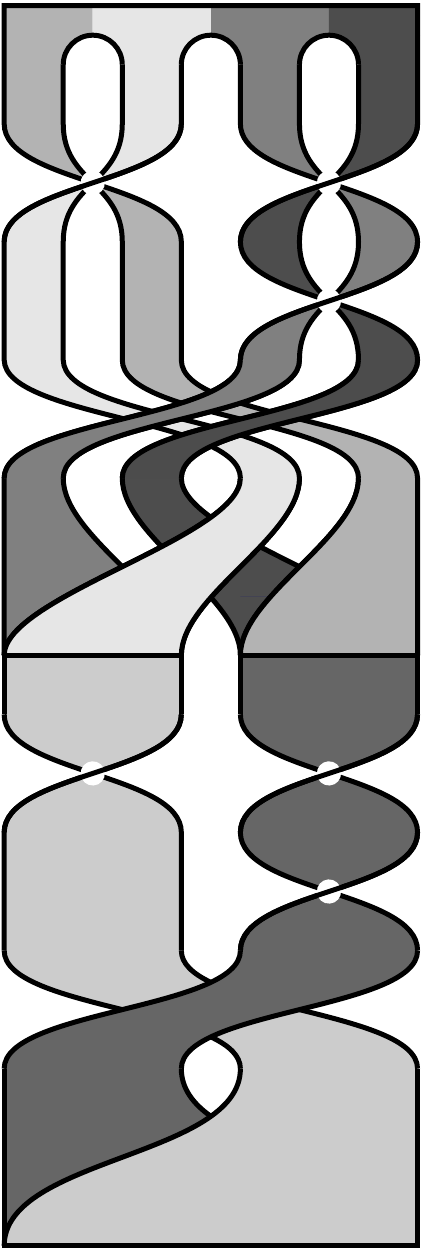}
    \put(-74,215){\small$u_2$}
    \put(-54,215){\small$u_1$}
    \put(-32,215){\color{white}\small$u_3$}
    \put(-12,215){\color{white}\small$u_4$}
    \put(-65,95){$t_1$}
    \put(-22,95){\color{white}$t_2$}
    & \quad &
    \includegraphics[width=0.168\textwidth]{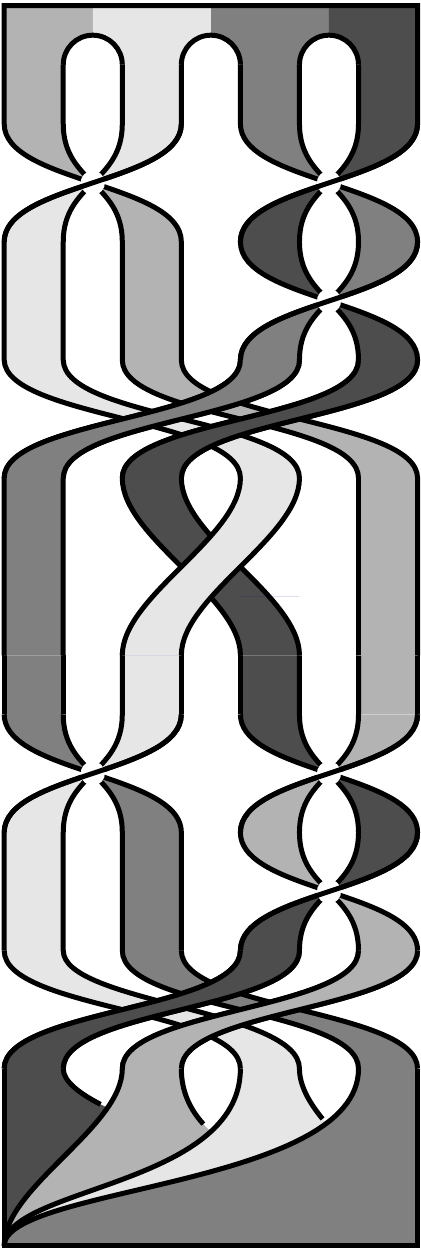}
    \put(-74,215){\small$u_2$}
    \put(-54,215){\small$u_1$}
    \put(-32,215){\color{white}\small$u_3$}
    \put(-12,215){\color{white}\small$u_4$}
    & \quad &
    \includegraphics[width=0.168\textwidth]{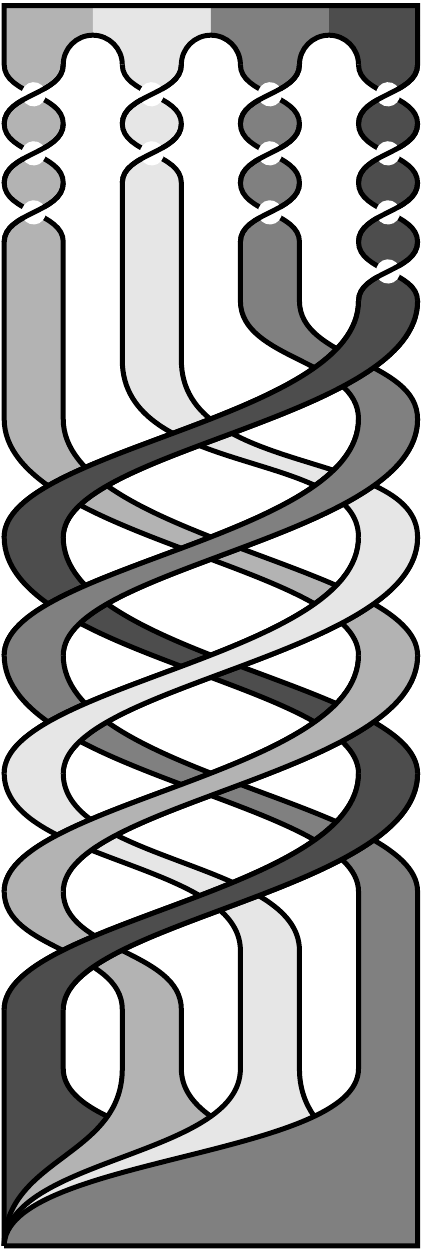}
    \put(-74,215){\small$u_2$}
    \put(-54,215){\small$u_1$}
    \put(-32,215){\color{white}\small$u_3$}
    \put(-12,215){\color{white}\small$u_4$}
  \end{tabular}
  \caption{
    Example \ref{exa:03}. (a) The two initial mixers. (b-c) The
    forward evolution of the strip of the resulting mixer $\mathcal{B}$.
    (d) The mixer of the symmetrical attractor.
  }
  \label{fig:example01}
  \vspace{-1em}
\end{figure}

\begin{example}
  \label{exa:03}
  In the article
  \cite{letellier1996evolution}, from an attractor solution to the
  Burke-Shaw system, the template of a symmetrical attractor is obtained and
  characterised by the linking matrix
  \begin{equation}
    \left[
      \begin{matrix}
      3 & 2 & 2 & 3 \\
      2 & 2 & 2 & 3 \\
      2 & 2 & 3 & 4 \\
      3 & 3 & 4 & 4
      \end{matrix}
    \right\rrbracket\;.
  \end{equation}
  Also, the image attractor mixer defined by the linking matrix
  \begin{equation}
    \left[
      \begin{matrix} 2 & 1 \\ 1 & 1
      \end{matrix}
    \right\rrbracket
  \end{equation}
  is obtained by the classical method \cite{letellier1995unstable}.  As shown
  in \cite{rosalie2013systematic}, the return map of the image attractor have
  to be built with a variable oriented from the inside to the outside to
  satisfy the conventions. This is not the case for the return map of the image
  system because $|x|$ is the chosen variable to build it (Fig.~10 of
  \cite{letellier1996evolution}). Thus, the order of the strips is reversed to get
  the exact mixer $\mathcal{A}$ of the image attractor defined by
  \begin{equation}
    A =
    \left[
      \begin{matrix} 1 & 1 \\ 1 & 2
      \end{matrix}
    \right\rrbracket\;.
  \end{equation}
  Then, the concatenation of $\mathcal{A}\oplus
  \mathcal{A}=\mathcal{B}$, with $\mathcal{B}$ a mixer defined by
  \begin{equation}
\resizebox{.9\textwidth}{!}{ $
    B=
    \begin{matrix}
      s_1 \\ s_2
    \end{matrix}
    \left[ \begin{matrix} 1 & 1 \\ 1 & 2
  \end{matrix}\right\rrbracket+
    \begin{matrix}
      t_1 \\ t_2
    \end{matrix}
  \left[ \begin{matrix} 1 & 1 \\ 1 & 2 \end{matrix}\right\rrbracket
  =
    \begin{matrix}
      u_2 \\ u_1 \\ u_3 \\ u_4
    \end{matrix}
  \left[\left[\begin{matrix}
      1 & 1 & 1 & 1 \\
      1 & 1 & 1 & 1 \\
      1 & 1 & 2 & 2 \\
      1 & 1 & 2 & 2
    \end{matrix}\right]
    +
    \left[\begin{matrix}
      0 & 0 & 0 & 0 \\
      0 & 0 & 0 & 1 \\
      0 & 0 & 0 & 0 \\
      0 & 1 & 0 & 0
  \end{matrix}\right]
  +
  \left[\begin{matrix}
    2 & 1 & 1 & 2 \\
    1 & 1 & 1 & 1 \\
    1 & 1 & 1 & 1 \\
    2 & 1 & 1 & 2
  \end{matrix}\right] \right\rrbracket
  =
    \begin{matrix}
      u_2 \\ u_1 \\ u_3 \\ u_4
    \end{matrix}
  \left[  \begin{matrix}
      3 & 2 & 2 & 3 \\
      2 & 2 & 2 & 3 \\
      2 & 2 & 3 & 3 \\
      3 & 3 & 3 & 4
    \end{matrix}\right\rrbracket
    \begin{matrix}
      u_2 \\ u_1 \\ u_3 \\ u_4
    \end{matrix} $}
    \label{eq:burkeshaw}
  \end{equation}
  Figure~\ref{fig:example01} shows the different steps of the Theorem
  \ref{theo:concatMixer} applied to this example. This concatenation permits
  to obtain the mixer of the symmetrical attractor that can be confirmed by
  the linking numbers between pairs of orbits \cite{lesceller1994algebraic}.
\end{example}

\subsection{Symmetry}
\label{sub:symmetry}

Finally, we also present the algebraical relation between two symmetric
templates \cite{rosalie2013systematic}. First, when two templates
$\mathcal{T}$ and $\mathcal{T}'$ are symmetric by reflection, the linking
matrix $T$ of $\mathcal{T}$ is the transposed linking matrix $T'$ of
$\mathcal{T}'$, thus, it is noted $T'=T^p$.

\begin{definition}\cite{rosalie2013systematic}
    Given two templates $\mathcal{T}$ and $\mathcal{T}'$ with respectively
    linking matrix $T$ and $T'$, if
\begin{equation}
    \resizebox{.25\textwidth}{!}{$
    T' = -
    \left[
    \begin{matrix}
        0 & 1 & \cdots & 1 \\
        1 & \ddots & \ddots & \vdots \\
        \vdots & \ddots & \ddots & 1 \\
        1 & \cdots & 1 &  0
    \end{matrix}
    \right] -T $}
    \label{eq:inversion}
\end{equation}
then $\mathcal{T}'$ is symmetric to $\mathcal{T}$ by inversion. It is noted $T' = \overline{T}$.
\end{definition}

As a consequence, using the previous relations, we can obtain the following
relations between the linking matrix of the templates of the Malasoma attractor
\eqref{eq:malasoma_v4}
\begin{equation}
    \label{eq:malasoma_01_reduced}
    L_a = L_b^{p} = \overline{L}_c = \overline{L}_d^{p} \, .
\end{equation}

\section{Elementary mixer}
\label{sec:elementary_matrix}

The \textit{direct template} of an attractor bounded by a genus--1 torus is an
ordered series of torsions and mixers.  The \textit{reduced template}
\cite{rosalie2015systematic} of an attractor is the concatenation of the
torsions and the mixers of a template; thus it is a mixer described by a
linking matrix. 

\begin{proposition}
    If there exists an algebraic relation between a minimal linking matrix and the linking matrix of a reduced template, then these matrices stand for the same dynamics up to torsions and symmetries.
\end{proposition}

\begin{proposition}
    A \emph{chaotic mechanism} of an attractor bounded by a genus--1 torus is described by a unique \emph{elementary mixer} defined by the \emph{minimal linking matrix}. 
    \label{def:chaotic_mechanism}
\end{proposition}

\begin{proposition}
    If reduced templates of attractors bounded by a genus--1 torus have the same elementary mixer, then, the attractors have the same \emph{chaotic mechanism}.
\end{proposition}

    From the set of equivalent mixers of size $n$, successive criteria are minimised to obtain the minimal linking matrix $M$ of a chaotic mechanism to obtain the sparsest linking matrix:
\begin{enumerate}
    \item Infinite norm: $||M||_\infty$
    \item Absolute value of the trace: $|\mathrm{Tr}(M)|$
    \item Inverse maximum index to have the least sparse lines on the latest possible lines: $\displaystyle I_{\max}^r(M) =  n + 1 - \max\!\left( \underset{i \in \{1, \dots, n\}}{\operatorname{Argmax}} \; \sum_{j=1}^{n} \left| M_{i,j} \right| \right)$
    \item Minimum index to have the sparsest lines on the first lines:\newline $\displaystyle  I_{\min}(M) = \min\!\left( \underset{i \in \{1, \dots, n\}}{\operatorname{Argmin}} \; \sum_{j=1}^{n} \left| M_{i,j} \right| \right)$
    \item Decreasing trace count $ d(M) =\displaystyle \sum_{i = 1}^{n-1} \textbf{1}(M_{i+1, i+1} - M_{i,i} = -1)$
\end{enumerate}
For instance, the last criterion helps to discriminate the following linking matrices
\begin{equation}
    \label{eq:criterion5}
N = 
    \left[\begin{matrix}
-2 & -2 & -1 &  0 &  1 &  1 &  1\\
-2 & -1 & -1 &  0 &  1 &  1 &  1\\
-1 & -1 &  0 &  0 &  1 &  1 &  1\\
 0 &  0 &  0 &  1 &  1 &  1 &  1\\
 1 &  1 &  1 &  1 &  2 &  1 &  1\\
 1 &  1 &  1 &  1 &  1 &  1 &  1\\
 1 &  1 &  1 &  1 &  1 &  1 &  2
    \end{matrix} \right\rrbracket
\text{ and } N^p =
    \left[\begin{matrix}
 2 & 1 & 1 & 1 & 1 & 1 & 1\\
 1 & 1 & 1 & 1 & 1 & 1 & 1\\
 1 & 1 & 2 & 1 & 1 & 1 & 1\\
 1 & 1 & 1 & 1 & 0 & 0 & 0\\
 1 & 1 & 1 & 0 & 0 &-1 &-1\\
 1 & 1 & 1 & 0 &-1 &-1 &-2\\
 1 & 1 & 1 & 0 &-1 &-2 &-2
    \end{matrix} \right\rrbracket
\end{equation}
where $||N||_\infty = ||N^p||_\infty = 8$, $|Tr(N)| = |Tr(N^p)| = 3$, $I_{\max}^r(N) =I_{\max}^r(N^p) = 1$, $I_{\min}(N) = I_{\min}(N^p) = 4$ and because $d(N) = 1$ and $d(N^p) = 5$. 
This criterion is not necessary to discriminate elementary linking matrix of size $3$ to $8$ except for that specific case illustrated on Eq.~\ref{eq:criterion5}. % over 116 linking matrices of size $7$ and none of the 337 elementary linking matrix of size $8$.

This minimisation process uses algebraic relations between linking matrices to discriminate candidates that are sorted accordingly to these metrics.
A function is defined to find out the minimal linking matrix in the neighbourhood of a given linking matrix $M$.
The neighbourhood is defined with $N$, a set of 12 mixers with the same chaotic mechanism of size $n$ described with $M$. $N = \left\{A \cup B \cup C \right\}$ where 
$A = \left\{M, M^p, \overline{M}, \overline{M}^p \right\}$ to have direct equivalent mixer, the following to have closer equivalent mixer with an additional torsion: positive
$B = \left\{ M + 1, (M + 1)^p, \overline{(M + 1)}, \overline{(M + 1)}^p \right\}$ and negative
$C = \left\{ (M -1), (M -1)^p, \overline{(M -1)}, \overline{(M -1)}^p \right\}$.
The minimal linking matrix of this set is a candidate to be the elementary linking matrix of the mixer $M$.
An algorithm is then defined to obtain the elementary linking matrix of a mixer.
Given a mixer $M_0$, the minimal linking matrix in his neighbourhood is computed: the candidate $M_1$.
If $M_1 = M_0$, then $M_0$ is the elementary mixer.
Else, by recurrence, a new candidate $M_{n+1}$ is computed from the previous candidate $M_n$ until $M_{n+1} = M_n$, giving $M_n$ as the elementary mixer.

\begin{example}
    \label{exa:04}
    From example \ref{exa:03}, the linking matrix $B$ has 11 linking matrices in his neighbourhood that can be sorted accordingly to criteria to elect the minimal linking matrix in this set (Tab.~\ref{tab:B_iter_1}).
    After the first iteration, the proposed minimal linking matrix is $B-1$. The process is repeated until it converges when the third iteration (Tab.~\ref{tab:B_iter_all}) gives $(B-3)^p$ as a candidate which is also the minimal linking matrix in its onw neighbourhood (Tab.~\ref{tab:B_iter_4}). 

    \begin{table}[hb]
        \caption{Iterations necessary to obtain the minimal linking matrix of $B$.}
        \label{tab:B_iter_all}
\noindent\resizebox{\textwidth}{!}{
        \renewcommand{\arraystretch}{1.3}
        $
        \begin{array}{c|c|c|c}
            & \text{Iteration 1} & \text{Iteration 2} & \text{Iteration 3}  \\
            \hline
 B = \left[  \begin{matrix}
      3 & 2 & 2 & 3 \\
      2 & 2 & 2 & 3 \\
      2 & 2 & 3 & 3 \\
      3 & 3 & 3 & 4
    \end{matrix}\right\rrbracket
 & 
 B-1 = \left[  \begin{matrix}
      2 & 1 & 1 & 2 \\
      1 & 1 & 1 & 2 \\
      1 & 1 & 2 & 2 \\
      2 & 2 & 2 & 3
    \end{matrix}\right\rrbracket
 &
 B-2 = \left[  \begin{matrix}
      1 & 0 & 0 & 1 \\
      0 & 0 & 0 & 1 \\
      0 & 0 & 1 & 1 \\
      1 & 1 & 1 & 2
    \end{matrix}\right\rrbracket
 &
 (B-3)^p = \left[  \begin{matrix}
      1 & 0 & 0 & 0 \\
      0 & 0 & -1 & -1 \\
      0 & -1 & -1 & -1 \\
      0 & -1 & -1 & 0
    \end{matrix}\right\rrbracket
    \end{array}
$}
\end{table}

\begin{table}[p]
    \centering
    \caption{Linking matrix in the neighbourhood of matrix $B$ (Example \ref{exa:03})}
    \label{tab:B_iter_1}
    \resizebox{!}{.5\textheight}{
        \renewcommand{\arraystretch}{1.3}
    \begin{tabular}{ccccccc}
        Operation & Linking matrix & $||M||_\infty$ & $|Tr(M)|$ & $I_{\max}^r(M) $ & $I_{\min}(M) $ & $d(M)$  \\ \hline
        $B-1$ & $\left[\begin{matrix} 2  &  1  &  1  &  2 \\ 1  &  1  &  1  &  2 \\ 1  &  1  &  2  &  2 \\ 2  &  2  &  2  &  3 \end{matrix} \right\rrbracket     $ &    9 &     8 &             1         & 2      &         1 \\
        $(B-1)^p$ & $\left[\begin{matrix} 3  &  2  &  2  &  2 \\ 2  &  2  &  1  &  1 \\ 2  &  1  &  1  &  1 \\ 2  &  1  &  1  &  2 \end{matrix} \right\rrbracket     $ &    9 &     8 &             4         & 3      &         2 \\
        $\overline{B - 1}$ & $\left[\begin{matrix} -2  & -2  & -2  & -3 \\ -2  & -1  & -2  & -3 \\ -2  & -2  & -2  & -3 \\ -3  & -3  & -3  & -3 \end{matrix} \right\rrbracket $ &   12 &     8 &             1         & 2      &         2 \\
        $\overline{B - 1}^p$ & $\left[\begin{matrix} -3  & -3  & -3  & -3 \\ -3  & -2  & -2  & -2 \\ -3  & -2  & -1  & -2 \\ -3  & -2  & -2  & -2 \end{matrix} \right\rrbracket $ &   12 &     8 &             4         & 3      &         1 \\
        $B$ & $\left[\begin{matrix} 3  &  2  &  2  &  3 \\ 2  &  2  &  2  &  3 \\ 2  &  2  &  3  &  3 \\ 3  &  3  &  3  &  4 \end{matrix} \right\rrbracket     $ &   13 &    12 &             1         & 2      &         1 \\
        $B^p$ & $\left[\begin{matrix} 4  &  3  &  3  &  3 \\ 3  &  3  &  2  &  2 \\ 3  &  2  &  2  &  2 \\ 3  &  2  &  2  &  3 \end{matrix} \right\rrbracket     $ &   13 &    12 &             4         & 3      &         2 \\
            $\overline{B}$ & $\left[\begin{matrix} -3  & -3  & -3  & -4 \\ -3  & -2  & -3  & -4 \\ -3  & -3  & -3  & -4 \\ -4  & -4  & -4  & -4 \end{matrix} \right\rrbracket $ &   16 &    12 &             1         & 2      &         2 \\
        $\overline{B}^p$ & $\left[\begin{matrix} -4  & -4  & -4  & -4 \\ -4  & -3  & -3  & -3 \\ -4  & -3  & -2  & -3 \\ -4  & -3  & -3  & -3 \end{matrix} \right\rrbracket $ &   16 &    12 &             4         & 3      &         1 \\
        $B + 1$ & $\left[\begin{matrix} 4  &  3  &  3  &  4 \\ 3  &  3  &  3  &  4 \\ 3  &  3  &  4  &  4 \\ 4  &  4  &  4  &  5 \end{matrix} \right\rrbracket     $ &   17 &    16 &             1         & 2      &         1 \\
        ${B+1}^p$ & $\left[\begin{matrix} 5  &  4  &  4  &  4 \\ 4  &  4  &  3  &  3 \\ 4  &  3  &  3  &  3 \\ 4  &  3  &  3  &  4 \end{matrix} \right\rrbracket     $ &   17 &    16 &             4         & 3      &         2 \\
        $\overline{B+1}$ & $\left[\begin{matrix} -4  & -4  & -4  & -5 \\ -4  & -3  & -4  & -5 \\ -4  & -4  & -4  & -5 \\ -5  & -5  & -5  & -5 \end{matrix} \right\rrbracket $ &   20 &    16 &             1         & 2      &         2 \\
        $\overline{B+1}^p$ & $\left[\begin{matrix} -5  & -5  & -5  & -5 \\ -5  & -4  & -4  & -4 \\ -5  & -4  & -3  & -4 \\ -5  & -4  & -4  & -4 \end{matrix} \right\rrbracket $ &   20 &    16 &             4         & 3      &         1 \\
\end{tabular}}
\end{table}

\begin{table}[p]
    \centering
    \caption{Linking matrix in the neighbourhood of matrix $C = (B-3)^p$.}
    \label{tab:B_iter_4}
    \resizebox{!}{.5\textheight}{
        \renewcommand{\arraystretch}{1.3}
    \begin{tabular}{ccccccc}
        Operation & Linking matrix & $||M||_\infty$ & $|Tr(M)|$ & $I_{\max}^r(M) $ & $I_{\min}(M) $ & $d(M)$  \\ \hline
$C$ & $\left[\begin{matrix} 1  &  0  &  0  &  0 \\ 0  &  0  & -1  & -1 \\ 0  & -1  & -1  & -1 \\ 0  & -1  & -1  &  0 \end{matrix} \right\rrbracket $     & 3   &  0   &          2  &       1&              2 \\
$C^p$ & $\left[\begin{matrix} 0  & -1  & -1  &  0 \\ -1  & -1  & -1  &  0 \\ -1  & -1  &  0  &  0 \\ 0  &  0  &  0  &  1 \end{matrix} \right\rrbracket $   & 3   &  0   &          3  &       4&              1 \\
    $\overline{C^p}$ & $\left[\begin{matrix} 0  &  0  &  0  & -1 \\ 0  &  1  &  0  & -1 \\ 0  &  0  &  0  & -1 \\ -1  & -1  & -1  & -1 \end{matrix} \right\rrbracket $    & 4   &  0   &          1  &       1&              2 \\
    $\overline{C} $ & $\left[\begin{matrix} -1  & -1  & -1  & -1 \\ -1  &  0  &  0  &  0 \\ -1  &  0  &  1  &  0 \\ -1  &  0  &  0  &  0 \end{matrix} \right\rrbracket $ & 4   &  0   &          4  &       2&              1 \\
    $\overline{C - 1}$ & $\left[\begin{matrix} 0  &  0  &  0  &  0 \\ 0  &  1  &  1  &  1 \\ 0  &  1  &  2  &  1 \\ 0  &  1  &  1  &  1 \end{matrix} \right\rrbracket $     & 4   &  4   &          2  &       1&              1 \\
 $\overline{C - 1}^p$  & $\left[\begin{matrix} 1  &  1  &  1  &  0 \\ 1  &  2  &  1  &  0 \\ 1  &  1  &  1  &  0 \\ 0  &  0  &  0  &  0 \end{matrix} \right\rrbracket $     & 4   &  4   &          3  &       4&              2 \\
     $(C+1)^p$ & $\left[\begin{matrix} 1  &  0  &  0  &  1 \\ 0  &  0  &  0  &  1 \\ 0  &  0  &  1  &  1 \\ 1  &  1  &  1  &  2 \end{matrix} \right\rrbracket $     & 5   &  4   &          1  &       2&              1 \\
$(C+1)$ & $\left[\begin{matrix} 2  &  1  &  1  &  1 \\ 1  &  1  &  0  &  0 \\ 1  &  0  &  0  &  0 \\ 1  &  0  &  0  &  1 \end{matrix} \right\rrbracket $     & 5   &  4   &          4  &       3&              2 \\
$(C - 1)$ & $\left[\begin{matrix} 0  & -1  & -1  & -1 \\ -1  & -1  & -2  & -2 \\ -1  & -2  & -2  & -2 \\ -1  & -2  & -2  & -1 \end{matrix} \right\rrbracket $  & 7   &  4   &          2  &       1&              2 \\
$(C - 1)^p$ & $\left[\begin{matrix} -1  & -2  & -2  & -1 \\ -2  & -2  & -2  & -1 \\ -2  & -2  & -1  & -1 \\ -1  & -1  & -1  &  0 \end{matrix} \right\rrbracket $ & 7   &  4   &          3  &       4&              1 \\
$\overline{(C + 1)}^p$ & $\left[\begin{matrix} -1  & -1  & -1  & -2 \\ -1  &  0  & -1  & -2 \\ -1  & -1  & -1  & -2 \\ -2  & -2  & -2  & -2 \end{matrix} \right\rrbracket $ & 8   &  4   &          1  &       2&              2 \\
$\overline{(C + 1)}$ & $\left[\begin{matrix} -2  & -2  & -2  & -2 \\ -2  & -1  & -1  & -1 \\ -2  & -1  &  0  & -1 \\ -2  & -1  & -1  & -1 \end{matrix} \right\rrbracket $ & 8   &  4   &          4  &       3&              1 \\
\end{tabular}}
\end{table}

Consequently, the minimal matrix of $B$ is defined by the following linking matrix 
\begin{equation}
 (B-3)^p = \left[  \begin{matrix}
      1 & 0 & 0 & 0 \\
      0 & 0 & -1 & -1 \\
      0 & -1 & -1 & -1 \\
      0 & -1 & -1 & 0
    \end{matrix}\right\rrbracket\;.
\end{equation}
Thus the elementary mixer of the Burke-Shaw attractor is defined by this minimal linking matrix. 
This minimal linking matrix is also describing the template of the Malasoma attractor ($L_a$ of Eq.~\ref{eq:malasoma_v4}).
We can deduce that the Malasoma attractor and the Burke-Shaw attractor have the same chaotic mechanism.

\end{example}

\section{Tables of chaotic mechanisms}
\label{sec:table_of_chaotic_mechanisms}

\subsection{Algorithm generating elementary linking matrices of chaotic mechanisms}
\label{sub:algorithm}

To generate all chaotic mechanisms, we propose an iterative process using concatenation. 
Concatenating two mixers having $n$ and $m$ strips gives a mixer of $m\times n$ strips. 
The algorithm starts with the unique elementary mixer with two strips defined by this elementary linking matrix:
\begin{equation}
    a = \left[\begin{matrix} 0 & 0 \\ 0 &
        1\end{matrix}\right\rrbracket\;.
\end{equation}
Concatenating this mixer with itself provides a mixer with four strips:
\begin{equation}
    \left[\begin{matrix} 0 & 0 \\ 0 & 1\end{matrix}\right\rrbracket +
    \left[\begin{matrix} 0 & 0 \\ 0 & 1\end{matrix}\right\rrbracket = 
        \left[\begin{matrix} 0 & 0 & 0 & 0\\
                0 & 1 & 1 & 1 \\
                0 & 1 & 2 & 1 \\
                0 & 1 & 1 & 1 \\
        \end{matrix}\right\rrbracket
\end{equation}
having two submixers of size 3 with their associated elementary mixer:
\begin{equation}
    \renewcommand{\arraystretch}{1.3}
    \begin{array}{c|c}
        \text{submixer} & \text{elementary mixer} \\ \hline
        \left[\begin{matrix} 0 & 0 & 0 \\
                0 & 1 & 1  \\
                0 & 1 & 2  \\
            \end{matrix}\right\rrbracket  &  
        \left[\begin{matrix} 1 & 0 & 0 \\
                0 & 0 & -1  \\
                0 & -1 & -1  \\
            \end{matrix}\right\rrbracket = c
            \\
        \left[\begin{matrix} 
                 1 & 1 & 1 \\
                 1 & 2 & 1 \\
                 1 & 1 & 1 \\
             \end{matrix}\right\rrbracket  &
        \left[\begin{matrix} 0 & 0 & 0 \\
                0 & 1 & 0  \\
                0 & 0 & 0  \\
            \end{matrix}\right\rrbracket = b
    \end{array}
\end{equation}

Thus, it produces a set of mixers of size 3 holding two elementary mixers $\{b, c\}$.
This process of concatenation is repeated with the mixers of size 3 creating the set: 
$\{b+b, b+c, c+b, c+c\}$ from which submixers of size 4 and their elementary mixers are extracted. 
It leads to a set of five elementary mixers of size 4.
By definition, concatenation allows to combine mechanisms one after the other. 
By concatenating of all the mechanisms, it generates all the possible mechanisms of size $n$ included in larger mixers because mixers of size $n-1$ are present with all possible additional strips on the left and on the right.
Concatenation achieves the extension of a mechanism within all other possible mechanisms.
The algorithm is thus the following one. 
Each pair of elementary mixers of size $n$ are concatenated to generated large mixers where submixers of size $n+1$ are extracted. 
The elementary mixers of these sumbmixers constitute the list of elementary mixers of size $n+1$ up to size equal to eight. 

\begin{table}[h]
    \centering
    \caption{Elementary mixers set size depending on the size of linking matrix.}
    \label{tab:size_of_set}
    \begin{tabular}{c|c}
        Size of the mixers & Size of the set of elementary mixers \\ \hline
        2 & 1  \\
        3 & 2  \\
        4 & 5  \\
        5 & 14 \\
        6 & 38 \\
        7 & 116\\
        8 & 337\\
    \end{tabular}
\end{table}

This algorithm forms sets of elementary mixers (Tab.~\ref{tab:size_of_set}).
We chose to generate mechanisms from concatenation that create complex and large mixers from which we take submixers.
Mart\'in and Used proposed a partial theoretical method to create new mixers and list possible templates \cite{martin2009simplified}.
Chaotic mechanism of the templates of \cite{martin2009simplified} are computed: in Fig.~2 of \cite{martin2009simplified}, the mechanisms $2A$ and $2C$ have the same chaotic mechanism whose elementary linking matrix is $c$ and the mechanisms $2B$ and $2D$ have the same chaotic mechanism whose elementary linking matrix is $b$.
Their description includes an additional vector to describe template that is not required for us because linking matrices respect the Tufillaro's convention \cite{tufillaro1990relative}.

\subsection{Chaotic mechanisms}
\label{sub:chaotic_mechanisms}

The nontrivial part of the topological characterisation method is to provide a
template (and its associated linking matrix) of the chaotic attractor.
This is mandatory to prove that the template reproduces the dynamics of the system by having the theoretical linking numbers of the template equal to the linking numbers numerically obtained (Sec.~\ref{sec:templates_of_chaotic_attractor}).
The construction of a template is generally made strip by strip, starting to find the relative organisation of two adjacent strips and adding a new strip to the left or to the right. 
Therefore, we organise the chaotic mechanism made of $n$ strips depending on
the chaotic mechanisms made of $n-1$ strips. 
This allows to easily link and build templates in regards of the two chaotic mechanisms made of $n-1$ strips they contain.
There is only one chaotic mechanism made of two strips (size 2), it is defined by the elementary mixer
\begin{equation}
    \text{a}=\left[\begin{matrix} 0 & 0 \\ 0 &
        1\end{matrix}\right\rrbracket\;,
\end{equation}
In addition to the elementary linking matrices and using \texttt{cate} software \cite{olszewski2018visualizing}, chaotic mechanisms are represented in the following tables:  
\begin{itemize}
    \item Tab.~\ref{tab:size3} presents mixers of size 3.
    \item Tab.~\ref{tab:size4} presents mixers of size 4.
    \item Tab.~\ref{tab:size5} presents mixers of size 5.
\end{itemize}
We restrict our study to chaotic mechanisms of size up to 5, as most chaotic attractors reported in the literature involve mechanisms of size 4 or smaller. 
The code used is provided (\ref{app:implementation}), however, one can generate elementary matrices of larger sizes when needed and can also extract the elementary matrix from any valid linking matrix (see \cite{olszewski2018visualizing} for the validity criterion).

\begin{table}[htpb]
    \centering
    \renewcommand{\arraystretch}{1.1}
    \caption{
        Two chaotic mechanisms with three strips : $\{b, c\}$.
    }
    \vspace{.2em}
    \label{tab:size3}
    \resizebox{.6\textwidth}{!}{
  % \begin{tabular}{*{2}{>{\centering}m{0.25\textwidth}}{>{}m{0.35\textwidth}}}
        \begin{tabular}{c?c}
        \begin{tabular}{lr} & Right elementary \\ \\ \\ Left elementary \end{tabular} &
         $a=\left[\begin{matrix} 0 & 0 \\ 0 &
             1\end{matrix}\right\rrbracket$ \includegraphics[width=6em,
             align=c,trim={-1em -1em -1em -1em}]{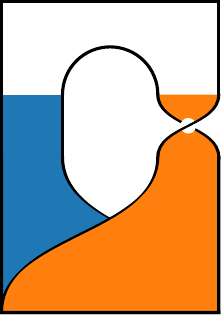} \\ \thickhline 
        $a=\left[\begin{matrix} 0 & 0 \\ 0 & 1\end{matrix}\right\rrbracket$ \includegraphics[width=6em,
             align=c,trim={-1em -1em -1em -1em}]{2x2_001.pdf} &
        $b=\left[\begin{matrix}  0 & 0 & 0 \\ 0 & 1 & 0 \\ 0 & 0 & 0 \end{matrix}\right\rrbracket \includegraphics[width=8em, align=c,trim={-1em -1em -1em -1em}]{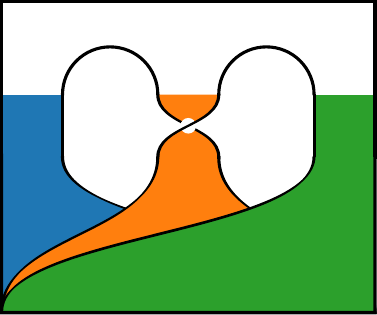} \quad
            c=\left[\begin{matrix}1 & 0 & 0 \\ 0 & 0 & -1 \\ 0 & -1 & -1
                \end{matrix}\right\rrbracket$ \includegraphics[width=8em,
             align=c,trim={-1em -1em -1em -1em}]{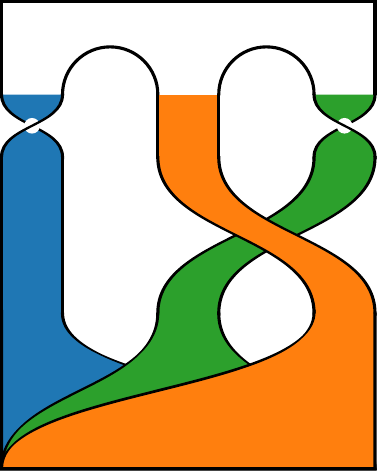}
        \end{tabular}}
\end{table}

\begin{table}[htpb]
    \centering
    \renewcommand{\arraystretch}{1.1}
    \caption{
        Five chaotic mechanisms with four strips : $\{d, e, f, g, h\}$.
    }
    \vspace{.2em}
    \label{tab:size4}
    \resizebox{\textwidth}{!}{
    \begin{tabular}{c?c|c}
        \begin{tabular}{lr} & Right elementary \\ \\ \\ \\ Left elementary \end{tabular} &
        $b=\left[\begin{matrix} 0 & 0 & 0 \\ 0 & 1 & 0 \\ 0 & 0 & 0
            \end{matrix}\right\rrbracket$\quad \includegraphics[width=8em, align=c,trim={-1em -1em -1em -1em}]{3x3_001.pdf} &
        $c=\left[\begin{matrix} 1 & 0 & 0 \\ 0 & 0 & -1 \\ 0 & -1 & -1 \end{matrix}\right\rrbracket$\quad\includegraphics[width=8em, align=c,trim={-1em -1em -1em -1em}]{3x3_002.pdf}
        \\ \thickhline
        $b=\left[\begin{matrix} 0 & 0 & 0 \\ 0 & 1 & 0 \\ 0 & 0 & 0
            \end{matrix}\right\rrbracket$ \quad\includegraphics[width=8em, align=c,trim={-1em -1em -1em -1em}]{3x3_001.pdf} &
        $d=\left[\begin{matrix}
            0 & 0 & 0 & 0 \\
            0 & 1 & 0 & 0 \\
            0 & 0 & 0 & 0 \\
            0 & 0 & 0 & 1 \\
        \end{matrix}\right\rrbracket$ \includegraphics[width=10em, align=c,trim={-1em -1em -1em -1em}]{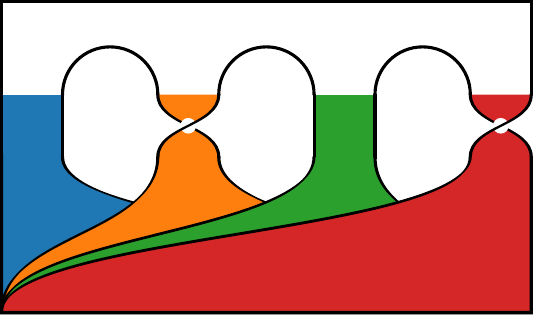}  &
        $e=\left[\begin{matrix}
          0  &  0  &  0  &  0 \\
          0  &  1  &  0  &  0 \\
          0  &  0  &  0  & -1 \\
          0  &  0  & -1  & -1
        \end{matrix}\right\rrbracket$ \includegraphics[width=10em, align=c,trim={-1em -1em -1em -1em}]{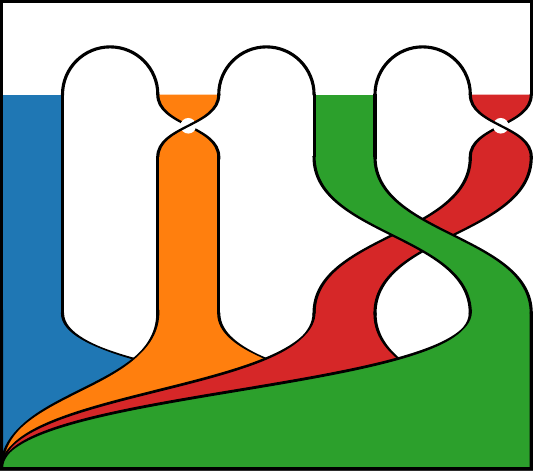} 
                \\ \hline
        $c=\left[\begin{matrix} 1 & 0 & 0 \\ 0 & 0 & -1 \\ 0 & -1 & -1
            \end{matrix}\right\rrbracket$ \includegraphics[width=6em, align=c]{3x3_002.pdf} &
$f=\left[\begin{matrix}
  1  &  0  &  0  &  0 \\
  0  &  0  & -1  & -1 \\
  0  & -1  & -1  & -1 \\
  0  & -1  & -1  &  0 
        \end{matrix}\right\rrbracket$\includegraphics[width=10em, align=c,trim={-1em -1em -1em -1em}]{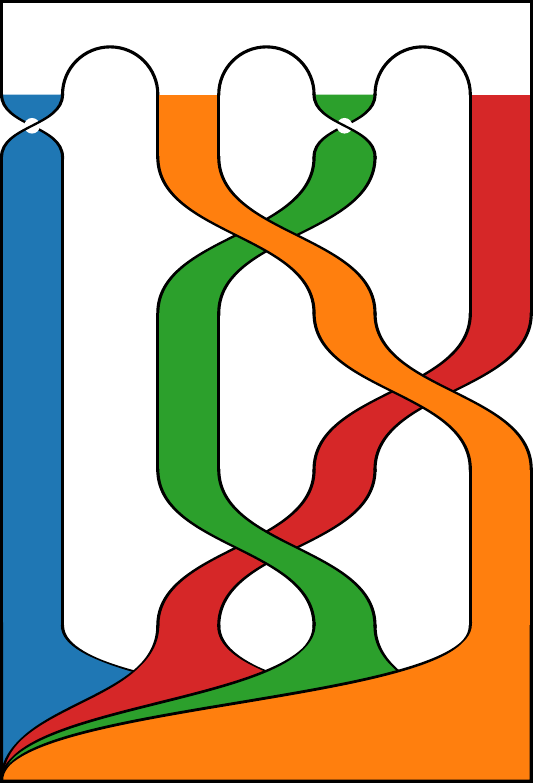}
&
        $g=\left[\begin{matrix}
            -1 & -1 & 0 & 0 \\
            -1 & 0 & 0 & 0 \\
            0 & 0 & 1 & 1 \\
            0 & 0 & 1 & 2 \\
        \end{matrix}\right\rrbracket$ \includegraphics[width=10em, align=c,trim={-1em -1em -1em -1em}]{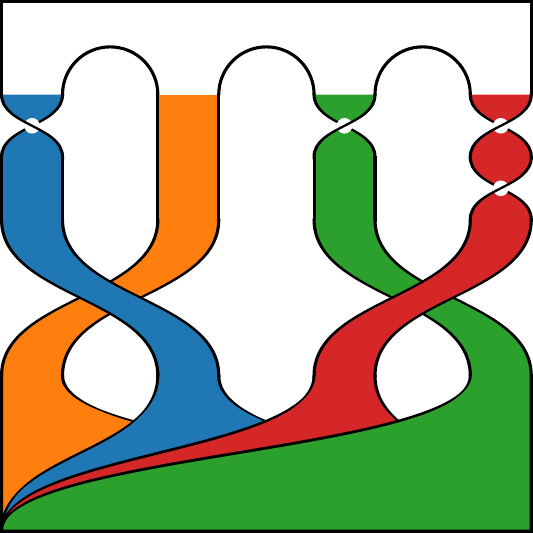}
        \quad
        $h=\left[\begin{matrix}
            -1 & -1 & -1 & -1 \\
            -1 & 0 & 0 & 0 \\
            -1 & 0 & 1 & 1 \\
            -1 & 0 & 1 & 2 \\
        \end{matrix}\right\rrbracket$\includegraphics[width=10em, align=c,trim={-1em -1em -1em -1em}]{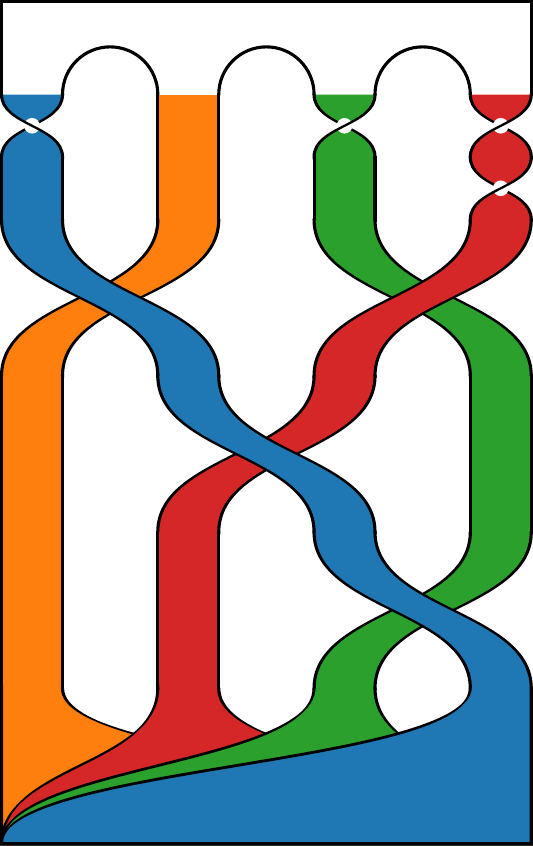}
\end{tabular}}
\end{table}

\begin{table}[htpb]
    \centering
    \renewcommand{\arraystretch}{1.2}
    \caption{
        Fourteen chaotic mechanisms with five strips : $\{i, \dots , v\}$.
    }
    \vspace{.2em}
    \label{tab:size5}
    \resizebox{1\textwidth}{!}{
    \begin{tabular}{c?c|c|c|c|c}
        \begin{tabular}{lr} & Right elementary \\ \\ \\ \\ Left elementary \end{tabular} &
        $d=\left[\begin{matrix}
            0 & 0 & 0 & 0 \\
            0 & 1 & 0 & 0 \\
            0 & 0 & 0 & 0 \\
            0 & 0 & 0 & 1 \\
        \end{matrix}\right\rrbracket$ &
        $e=\left[\begin{matrix}
          0  &  0  &  0  &  0 \\
          0  &  1  &  0  &  0 \\
          0  &  0  &  0  & -1 \\
          0  &  0  & -1  & -1
        \end{matrix}\right\rrbracket$ &
        $f=\left[\begin{matrix}
          1  &  0  &  0  &  0 \\
          0  &  0  & -1  & -1 \\
          0  & -1  & -1  & -1 \\
          0  & -1  & -1  &  0
        \end{matrix}\right\rrbracket$ &
        $g=\left[\begin{matrix}
            -1 & -1 & 0 & 0 \\
            -1 & 0 & 0 & 0 \\
            0 & 0 & 1 & 1 \\
            0 & 0 & 1 & 2 \\
        \end{matrix}\right\rrbracket$ &
        $h=\left[\begin{matrix}
            -1 & -1 & -1 & -1 \\
            -1 & 0 & 0 & 0 \\
            -1 & 0 & 1 & 1 \\
            -1 & 0 & 1 & 2 \\
        \end{matrix}\right\rrbracket$ \\ \thickhline
        $d=\left[\begin{matrix}
            0 & 0 & 0 & 0 \\
            0 & 1 & 0 & 0 \\
            0 & 0 & 0 & 0 \\
            0 & 0 & 0 & 1 \\
        \end{matrix}\right\rrbracket$ &
        $i=\left[\begin{matrix}
            0 & 0 & 0 & 0 & 0 \\
            0 & 1 & 0 & 0 & 0 \\
            0 & 0 & 0 & 0 & 0 \\
            0 & 0 & 0 & 1 & 0 \\
            0 & 0 & 0 & 0 & 0 \\
        \end{matrix}\right\rrbracket$
        & 
        $j=\left[\begin{matrix}
  1  &  0  &  0  &  0  &  0 \\
  0  &  0  &  0  &  0  &  0 \\
  0  &  0  &  1  &  0  &  0 \\
  0  &  0  &  0  &  0  & -1 \\
  0  &  0  &  0  & -1  & -1 
        \end{matrix}\right\rrbracket$
        &
        $k=\left[\begin{matrix}
  1  &  0  &  0  &  0  &  0 \\
  0  &  0  &  0  &  0  & -1 \\
  0  &  0  &  1  &  0  & -1 \\
  0  &  0  &  0  &  0  & -1 \\
  0  & -1  & -1  & -1  & -1 
        \end{matrix}\right\rrbracket$
        &
        \\ \hline
        $e=\left[\begin{matrix}
          0  &  0  &  0  &  0 \\
          0  &  1  &  0  &  0 \\
          0  &  0  &  0  & -1 \\
          0  &  0  & -1  & -1
        \end{matrix}\right\rrbracket$ &
                &
        $l =\left[\begin{matrix}
             -1  & -1  &  0  &  0  &  0 \\
             -1  &  0  &  0  &  0  &  0 \\
              0  &  0  &  1  &  0  &  0 \\
              0  &  0  &  0  &  0  & -1 \\
              0  &  0  &  0  & -1  & -1
        \end{matrix}\right\rrbracket$ &
        $m=\left[\begin{matrix}
                  0  &  0  &  0  &  0  &  0 \\
                  0  &  1  &  0  &  0  &  0 \\
                  0  &  0  &  0  & -1  & -1 \\
                  0  &  0  & -1  & -1  & -1 \\
                  0  &  0  & -1  & -1  &  0
        \end{matrix}\right\rrbracket
        $
        &
        &
        $
        n =\left[\begin{matrix}
                  0  &  0  &  0  &  0  &  0 \\
                  0  &  1  &  0  &  0  &  0 \\
                  0  &  0  &  0  & -1  & -1 \\
                  0  &  0  & -1  & -1  & -2 \\
                  0  &  0  & -1  & -2  & -2
        \end{matrix}\right\rrbracket
        $
        \\ \hline
        $f=\left[\begin{matrix}
             1  &  0  &  0  &  0 \\
             0  &  0  & -1  & -1 \\
             0  & -1  & -1  & -1 \\
             0  & -1  & -1  &  0
        \end{matrix}\right\rrbracket$ &
        $o =\left[\begin{matrix}
                  1  &  0  &  0  &  0  &  0 \\
                  0  &  0  & -1  & -1  & -1 \\
                  0  & -1  & -1  & -1  & -1 \\
                  0  & -1  & -1  &  0  & -1 \\
                  0  & -1  & -1  & -1  & -1
        \end{matrix}\right\rrbracket$
        &
        $p=\left[\begin{matrix}
                  1  &  0  &  0  &  0  &  0 \\
                  0  &  0  & -1  & -1  & -1 \\
                  0  & -1  & -1  & -1  & -1 \\
                  0  & -1  & -1  &  0  &  0 \\
                  0  & -1  & -1  &  0  &  1 
        \end{matrix}\right\rrbracket$
        &
        $q=\left[\begin{matrix}
              1  &  0  &  0  &  0  &  0 \\
              0  &  0  & -1  & -1  &  0 \\
              0  & -1  & -1  & -1  &  0 \\
              0  & -1  & -1  &  0  &  0 \\
              0  &  0  &  0  &  0  &  1 
        \end{matrix}\right\rrbracket$
        & 
        $\begin{matrix}
            r =\left[\begin{matrix}
                  0  & -1  & -1  &  0  &  0 \\
                 -1  & -1  & -1  &  0  &  0 \\
                 -1  & -1  &  0  &  0  &  0 \\
                  0  &  0  &  0  &  1  &  1 \\
                  0  &  0  &  0  &  1  &  2 
        \end{matrix}\right\rrbracket &
s = \left[\begin{matrix}
  0  & -1  & -1  &  0  &  1 \\
 -1  & -1  & -1  &  0  &  0 \\
 -1  & -1  &  0  &  0  &  0 \\
  0  &  0  &  0  &  1  &  1 \\
  1  &  0  &  0  &  1  &  2 
        \end{matrix}\right\rrbracket 
        \end{matrix}$ 
        &
        $t =\left[\begin{matrix}
                  1  &  1  &  1  &  0  & -1 \\
                    1  &  2  &  1  &  0  & -1 \\
                    1  &  1  &  1  &  0  & -1 \\
                    0  &  0  &  0  &  0  & -1 \\
                    -1  & -1  & -1  & -1  & -1
        \end{matrix}\right\rrbracket$
        \\ \hline
        $g=\left[\begin{matrix}
            -1 & -1 & 0 & 0 \\
            -1 & 0 & 0 & 0 \\
            0 & 0 & 1 & 1 \\
            0 & 0 & 1 & 2 \\
        \end{matrix}\right\rrbracket$
        & & & & &
        $u=\left[\begin{matrix}
  2  &  1  &  0  &  0  &  0 \\
  1  &  1  &  0  &  0  &  0 \\
  0  &  0  &  0  & -1  & -1 \\
  0  &  0  & -1  & -1  & -2 \\
  0  &  0  & -1  & -2  & -2 
        \end{matrix}\right\rrbracket$
        \\ \hline
        $h=\left[\begin{matrix}
            -1 & -1 & -1 & -1 \\
            -1 & 0 & 0 & 0 \\
            -1 & 0 & 1 & 1 \\
            -1 & 0 & 1 & 2 \\
        \end{matrix}\right\rrbracket$
        & & & & &
        $v=\left[\begin{matrix}
  2  &  1  &  1  &  1  &  1 \\
  1  &  1  &  0  &  0  &  0 \\
  1  &  0  &  0  & -1  & -1 \\
  1  &  0  & -1  & -1  & -2 \\
  1  &  0  & -1  & -2  & -2 
        \end{matrix}\right\rrbracket$ \\ \thickhline \\ \thickhline
    % \end{tabular}
% }
% \medskip
%
% \end{table}
% 
% 
% \begin{table}[htpb]
%     \centering
%     \renewcommand{\arraystretch}{1.1}
%     \caption{
%         Mixers of the fourteen chaotic mechanisms with five strips : $\{i, \dots , v\}$.
%     }
%     \vspace{.2em}
%     \label{tab:size5b}
    % \resizebox{1\textwidth}{!}{
    % \begin{tabular}{c?c|c|c|c|c}
        \begin{tabular}{lr} & Right elementary \\ \\ \\ \\ \\  \\ \\ \\ Left elementary \end{tabular} &
        $d$ \includegraphics[width=10em, align=c, trim={-1em -1em -1em -1em}]{4x4_001.pdf} &
        $e$ \includegraphics[width=10em, align=c, trim={-1em -1em -1em -1em}]{4x4_003.pdf} &
        $f$ \includegraphics[width=10em, align=c, trim={-1em -1em -1em -1em}]{4x4_002.pdf} &
        $g$ \includegraphics[width=10em, align=c, trim={-1em -1em -1em -1em}]{4x4_004.pdf} &
        $h$ \includegraphics[width=10em, align=c, trim={-1em -1em -1em -1em}]{4x4_005.pdf}
        \\ \thickhline
        $d$ \includegraphics[width=10em, align=c, trim={-1em -1em -1em -1em}]{4x4_001.pdf} &
        $i$ \includegraphics[width=12em, align=c, trim={-1em -1em -1em -1em}]{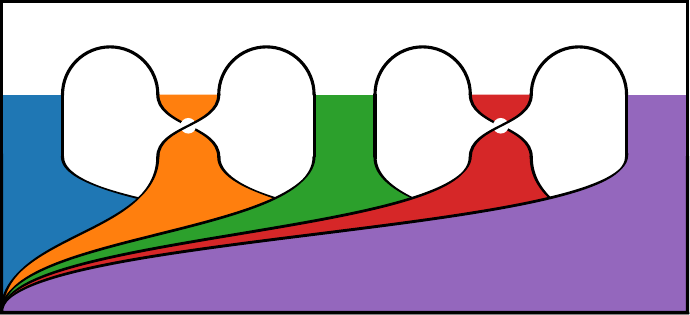} & 
        $j$ \includegraphics[width=12em, align=c, trim={-1em -1em -1em -1em}]{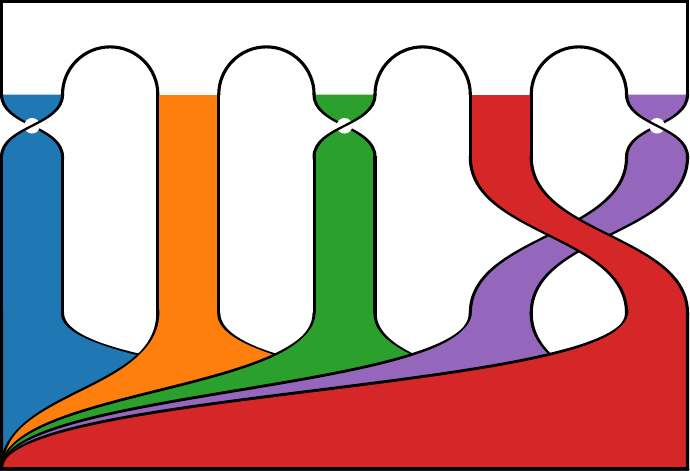} &
        $k$ \includegraphics[width=12em, align=c, trim={-1em -1em -1em -1em}]{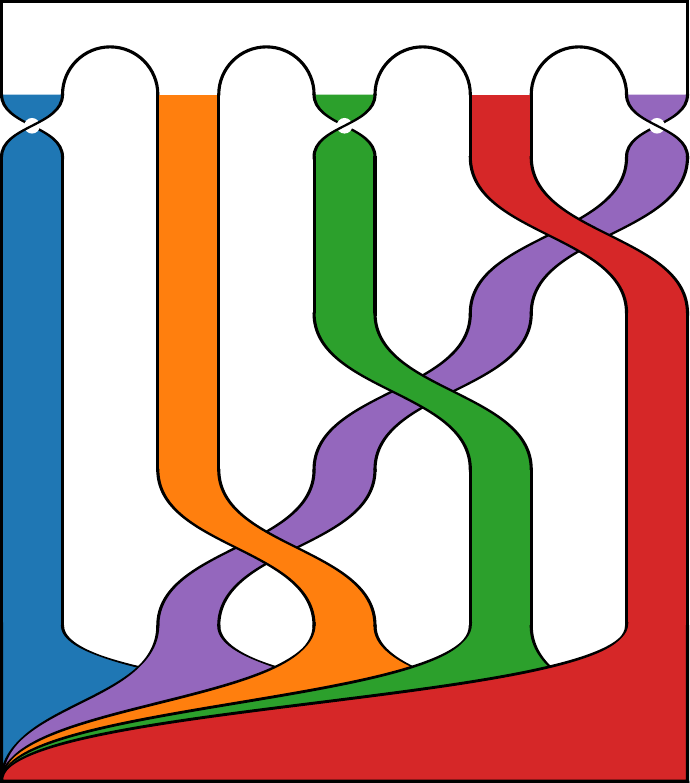}
        &
        \\ \hline
        $e$ \includegraphics[width=10em, align=c, trim={-1em -1em -1em -1em}]{4x4_003.pdf}&
        &
        $l$ \includegraphics[width=12em, align=c, trim={-1em -1em -1em -1em}]{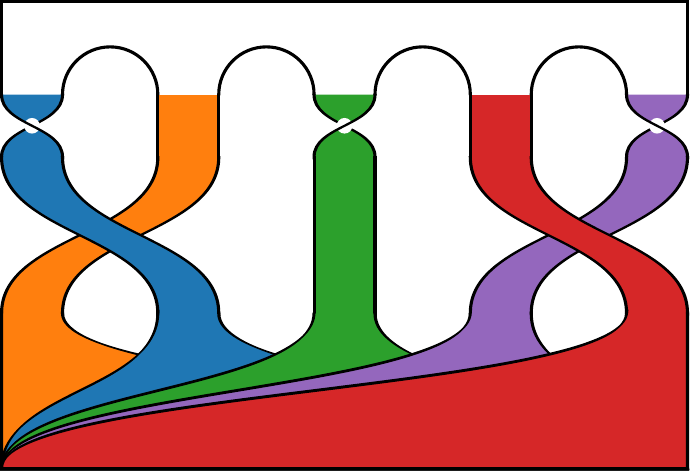}
        &
        $m$ \includegraphics[width=10em, align=c, trim={-1em -1em -1em -1em}]{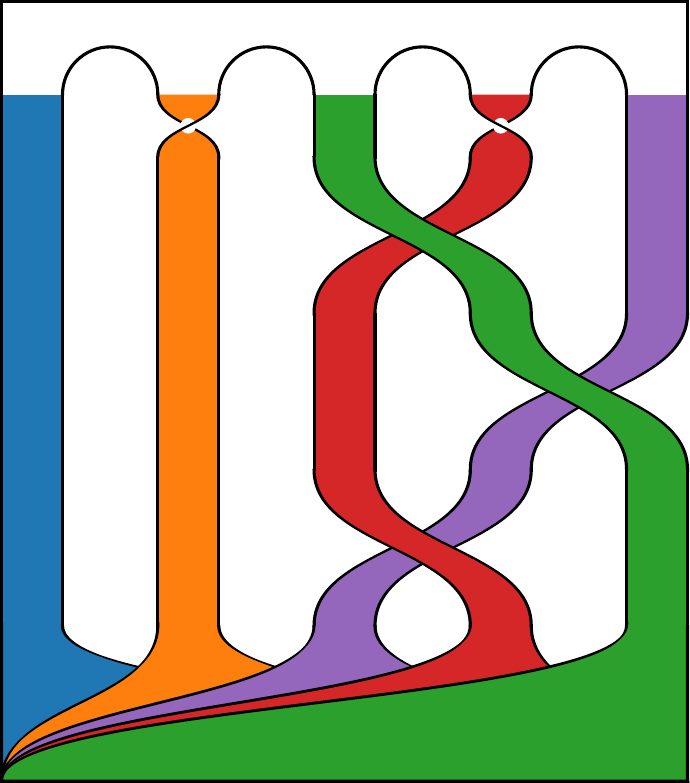}&
                                                                                          &
        $n$ \includegraphics[width=12em, align=c, trim={-1em -1em -1em -1em}]{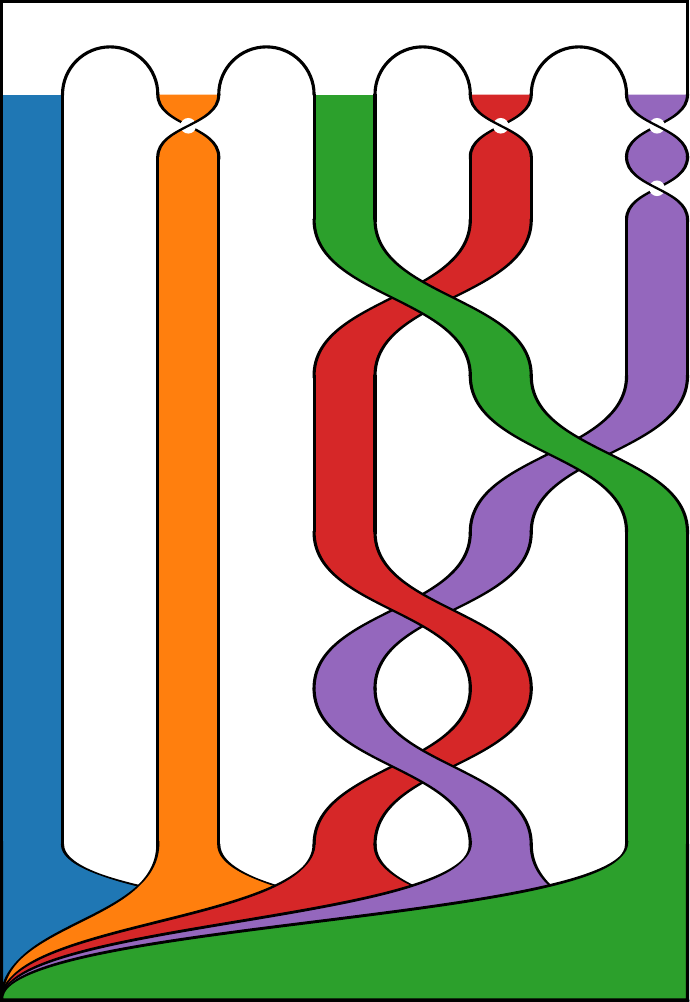}
        \\ \hline
        $f$ \includegraphics[width=10em, align=c, trim={-1em -1em -1em -1em}]{4x4_002.pdf}&
        $o$ \includegraphics[width=10em, align=c, trim={-1em -1em -1em -1em}]{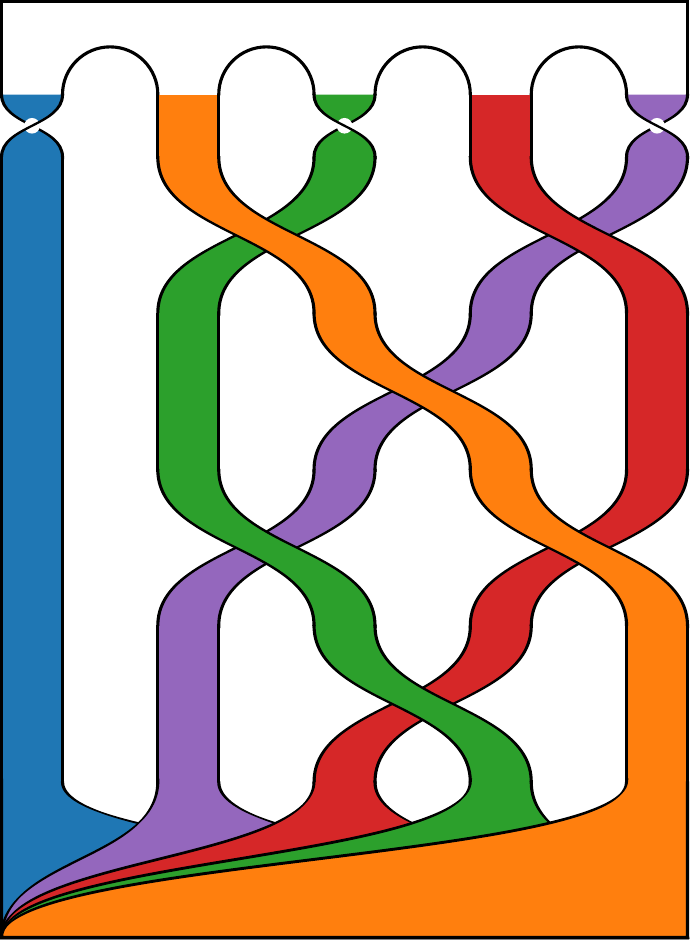}
        &
        $p$ \includegraphics[width=10em, align=c, trim={-1em -1em -1em -1em}]{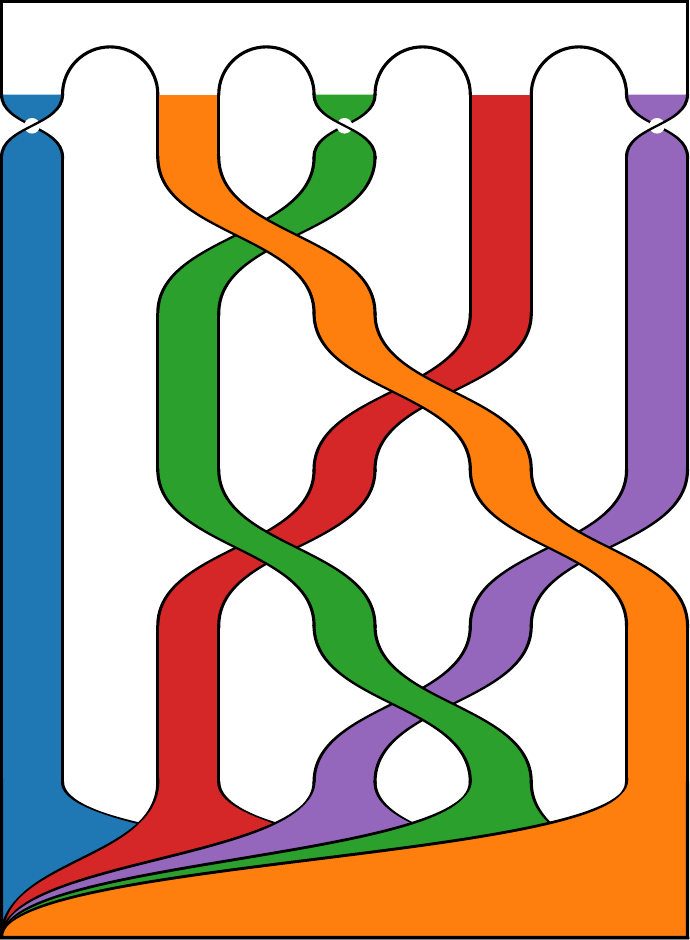}
        &
        $q$ \includegraphics[width=12em, align=c, trim={-1em -1em -1em -1em}]{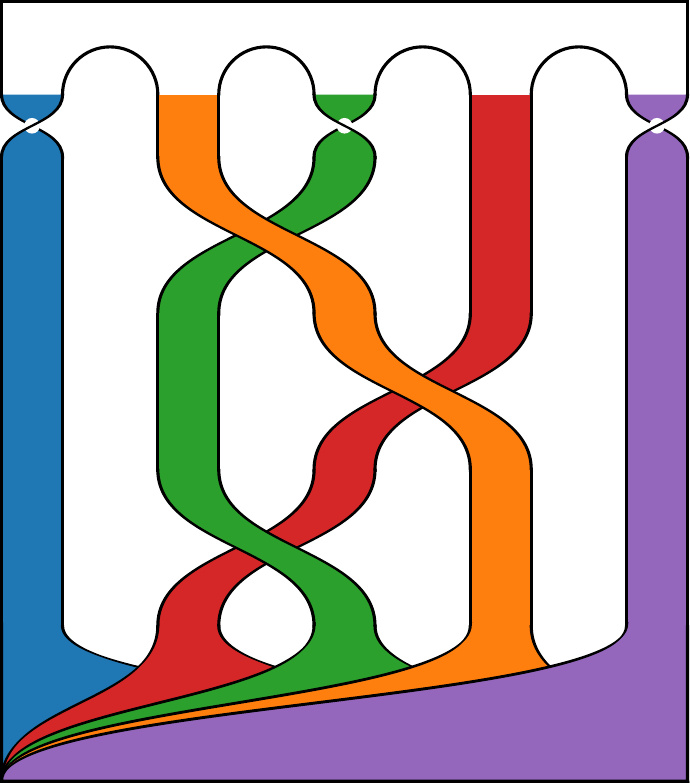} 
        &
        \begin{tabular}{cc}
            $r$ \includegraphics[width=12em, align=c, trim={-1em -1em -1em -1em}]{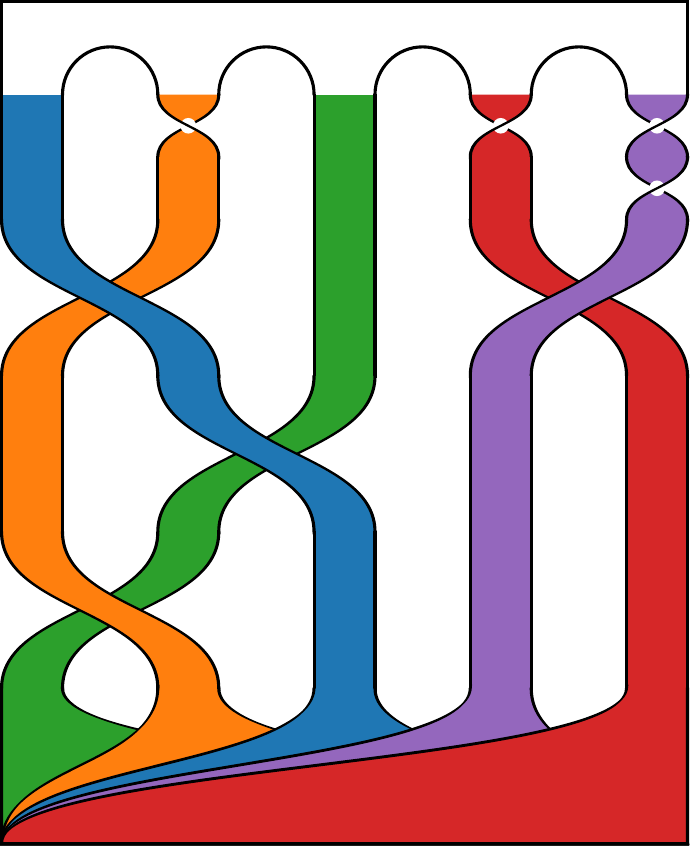} & 
            $s$ \includegraphics[width=12em, align=c, trim={-1em -1em -1em -1em}]{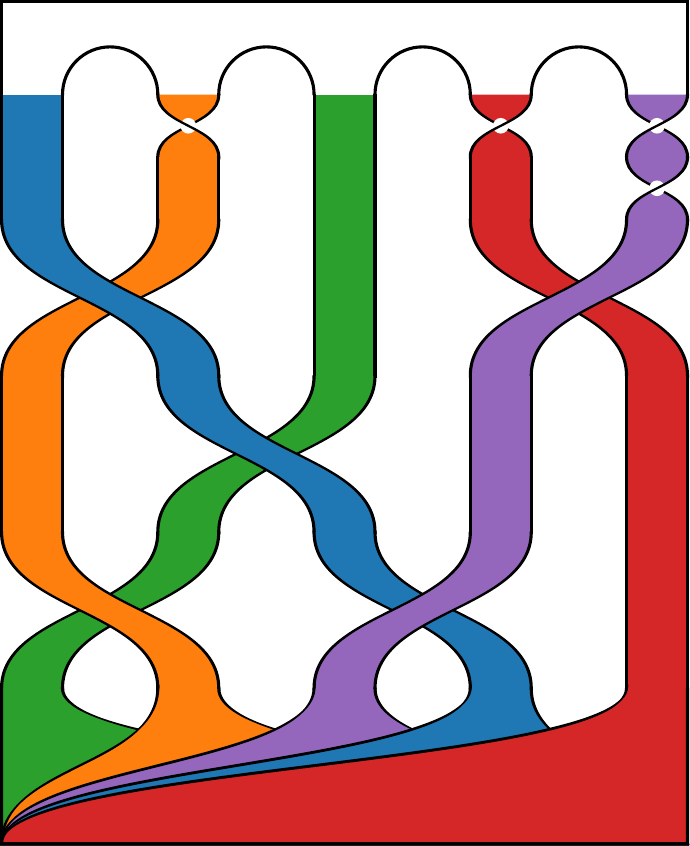}
        \end{tabular}
        &
        $t$ \includegraphics[width=10em, align=c, trim={-1em -1em -1em -1em}]{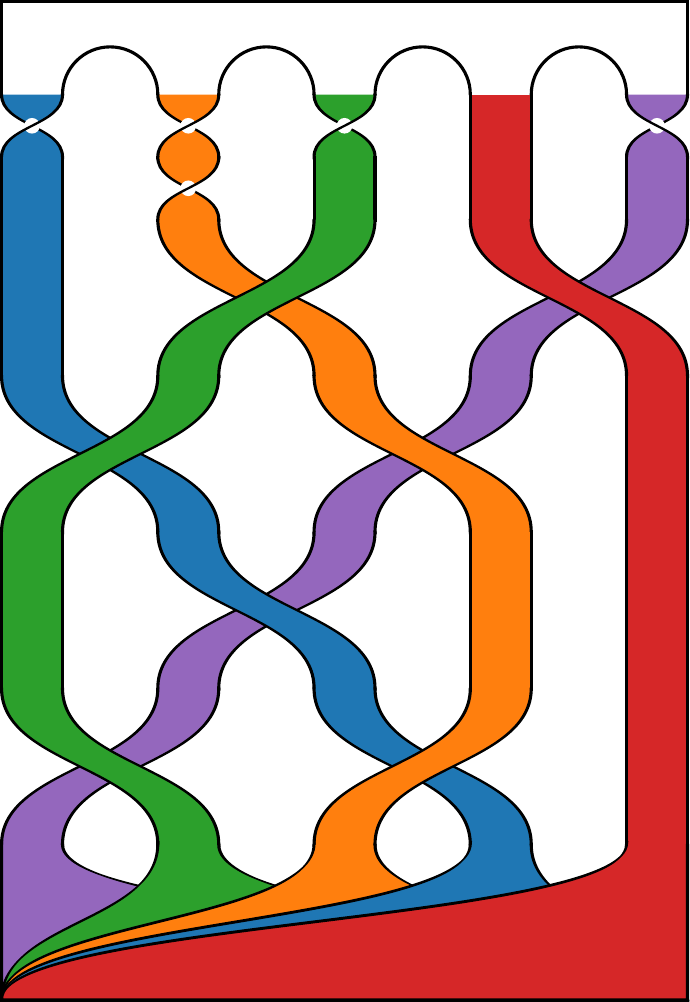}
        \\ \hline
        $g$ \includegraphics[width=10em, align=c, trim={-1em -1em -1em -1em}]{4x4_004.pdf}&
                                                                                          & & & & 
        $u$ \includegraphics[width=10em, align=c, trim={-1em -1em -1em -1em}]{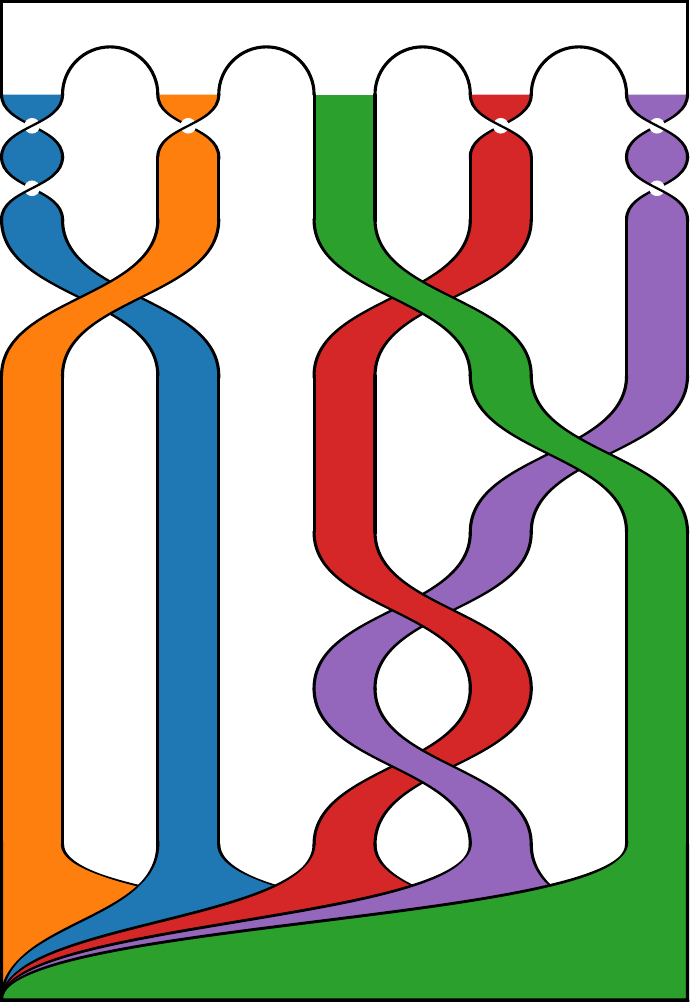}
        \\ \hline
        $h$ \includegraphics[width=10em, align=c, trim={-1em -1em -1em -1em}]{4x4_005.pdf}
        & &
        & &
        &
        $v$ \includegraphics[width=10em, align=c, trim={-1em -1em -1em -1em}]{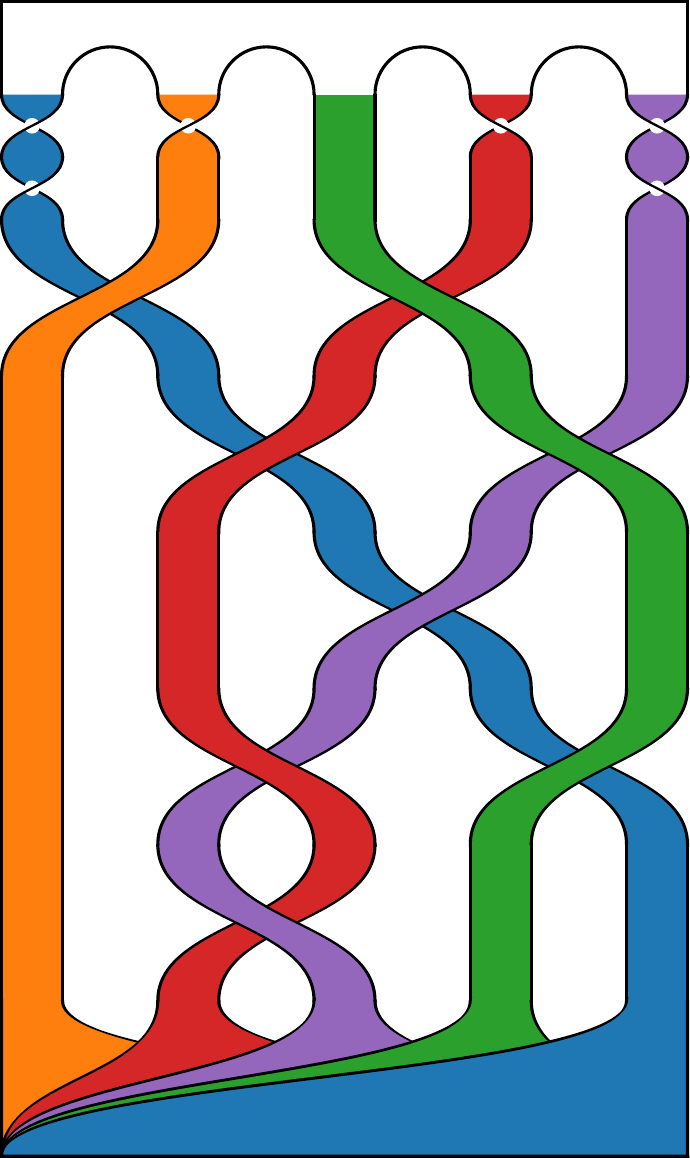}
    \end{tabular}
}
\end{table}

\subsection{Acceleration of the generation of all chaotic mechanisms}

\begin{proposition}
    Extending the chaotic mechanisms of size $n$ by adding a left and right fold is sufficient to generate all possible combinations of mechanisms of size $n+1$.
\end{proposition}

Adding a left (or right) fold to a mixer of size $n$ could be done with concatenation. 
The mixer after is inserted in both branches of the first mixer by definition of concatenation.
Concatenating a mixer of size $n$ after the elementary mixer $a$ is equivalent to place them side by side and connect them, then stretch this link so that the second mixer is above the other, and reciprocally, below, with $a^p$.
Doing so, it extends the mixer $n$ with a branch up or down. 
This is the case in Fig.~\ref{fig:example01} where $u_3$ extend the mixer made of $t_1$ and $t_2$ that is inserted in $s_1$. It results in the three first lines and row of the of $B$ \eqref{eq:burkeshaw}.

\begin{example}
    From the two mixers of size $3$ ($b$ and $c$), we summarise the results of the two method proposed to generate the list of chaotic mechanisms. Operations with \emph{full concatenation} of mixers generate four mixers of size $3 \times 3 = 9$ from which we can extract six mixers of size 4 to a total of 24. The \emph{extension by concatenation} reduces the number of mixers to test to three for each operation, the total is thus equal to 12.

    \begin{tabular}{cl}
        \multicolumn{2}{l}{\textbf{Full concatenation}} \\
        Operation & Elementary mixers \\ \hline
        $b + b$ & $d, e, f$ \\
        $b + c$ & $e, f, g, h$ \\
        $c + b$ & $d, e, f$ \\
        $c + c$ & $e, f, g, h$
    \end{tabular}
    \qquad
    \begin{tabular}{cl}
        \multicolumn{2}{l}{\textbf{Extension by concatenation}} \\
        Operation & Elementary mixers \\ \hline
        $a + b$ & $d, e, f$ \\
        $a^p + b$ & $d, e, f$ \\
        $a + c$ & $f, h$ \\
        $a^p + c$ & $e, f, g$ \\
    \end{tabular}
\end{example}

As illustrated in the example, the extension by concatenation reduces the number of mixers from which the elementary mixer must be extracted. 
The generation of all chaotic mechanisms can be accelerated: \emph{concatenation by extension} runs in $\mathcal{O}(n)$, whereas \emph{full concatenation} requires $\mathcal{O}(n^2)$.
Implementation (\ref{app:implementation}) of this algorithm confirm this acceleration with a significant decrease of computation time required to generate the same list of mixers for size below 9.

\section{Open problem}
We computed linking matrices of elementary mixer up to size 8 with criteria discriminating them. 
The fifth criterion helps to discriminate two symmetrical linking matrices by permutation \eqref{eq:criterion5}.
This is sufficient because examples of templates with more than eight branches are rare.
However, this fifth criterion is not enough to address the uniqueness of elementary linking matrices of size 9 and more. It would be ideal to have a more robust criterion.
\textit{Would it be possible to find a fifth criterion that would select a unique elementary matrix regardless of the size of the matrix?}
Examples of the three linking matrices of size 9 are given in Tab.~\ref{tab:size9} to have an overview of the problem (1024 elementary linking matrices of size 9 are obtained with the algorithm).
If one could be tempt to use only positive numbers into linking matrices of elementary mixers, it will first, lenghten templates (see Fig.~\ref{fig:example01}) and second, it will remove symmetrical structures that are inherent part of chaotic attractors \cite{Gilmore2007}.

\begin{table}[htbp]
    \caption{Examples of matrices not discriminated by the 5th criterion.}
    \resizebox{\textwidth}{!}{
    \begin{tabular}{ccc}
        Candidate \#1 & Candidate \#2 & Criteria\\  \hline  \\
        $\left[\begin{matrix}
                -1 &-1 &-1 &-1 &-1 &-1 &-1 &-1 &-1\\
                -1 & 0 & 0 & 0 & 0 & 0 & 0 & 0 &-1\\
                -1 & 0 & 1 & 1 & 1 & 1 & 1 & 0 &-1\\
                -1 & 0 & 1 & 2 & 1 & 1 & 1 & 0 &-1\\
                -1 & 0 & 1 & 1 & 1 & 0 & 0 & 0 &-1\\
                -1 & 0 & 1 & 1 & 0 & 0 & 0 & 0 &-1\\
                -1 & 0 & 1 & 1 & 0 & 0 & 1 & 0 &-1\\
                -1 & 0 & 0 & 0 & 0 & 0 & 0 & 0 &-1\\
                -1 &-1 &-1 &-1 &-1 &-1 &-1 &-1 &-1
            \end{matrix}\right\rrbracket$ & 
            $\left[\begin{matrix}
                -1  &-1 & -1 & -1 & -1 & -1 & -1  &-1 &-1\\
                -1  & 0 &  0 &  0 &  0 &  0 &  0  & 0 &-1\\
                -1  & 0 &  1 &  0 &  0 &  1 &  1  & 0 &-1\\
                -1  & 0 &  0 &  0 &  0 &  1 &  1  & 0 &-1\\
                -1  & 0 &  0 &  0 &  1 &  1 &  1  & 0 &-1\\
                -1  & 0 &  1 &  1 &  1 &  2 &  1  & 0 &-1\\
                -1  & 0 &  1 &  1 &  1 &  1 &  1  & 0 &-1\\
                -1  & 0 &  0 &  0 &  0 &  0 &  0  & 0 &-1\\
                -1  &-1 & -1 & -1 & -1 & -1 & -1  &-1 &-1\\
            \end{matrix}\right\rrbracket$ 
                      & $\begin{pmatrix}
                          9 \\ 3 \\ 1 \\ 2 \\4
                      \end{pmatrix}$ \\ \\
        $\left[\begin{matrix}
                    2 &  1 &  0 &  0 &  1 &  1 &  0 & -1 & -1 \\
                    1 &  1 &  0 &  0 &  0 &  0 &  0 & -1 & -1 \\
                    0 &  0 &  0 & -1 & -1 & -1 & -1 & -1 & -1 \\
                    0 &  0 & -1 & -1 & -1 & -1 & -1 & -1 & -1 \\
                    1 &  0 & -1 & -1 &  0 &  0 &  0 & -1 & -1 \\
                    1 &  0 & -1 & -1 &  0 &  1 &  0 & -1 & -1 \\
                    0 &  0 & -1 & -1 &  0 &  0 &  0 & -1 & -1 \\
                    1 & -1 & -1 & -1 & -1 & -1 & -1 & -1 & -2 \\
                    1 & -1 & -1 & -1 & -1 & -1 & -1 & -2 & -2 \\
            \end{matrix}\right\rrbracket$ & 
            $\left[\begin{matrix}
                    2 &  1 &  0 &  0 &  0 &  0 &  0 &  0 &  0 \\
                    1 &  1 &  0 &  0 &  0 &  0 &  0 &  0 &  0 \\
                    0 &  0 &  0 & -1 & -1 &  0 &  0 & -1 & -1 \\
                    0 &  0 & -1 & -1 & -1 &  0 &  0 & -1 & -2 \\
                    0 &  0 & -1 & -1 &  0 &  0 &  0 & -1 & -2 \\
                    0 &  0 &  0 &  0 &  0 &  1 &  0 & -1 & -1 \\
                    0 &  0 &  0 &  0 &  0 &  0 &  0 & -1 & -1 \\
                    0 &  0 & -1 & -1 & -1 & -1 & -1 & -1 & -2 \\
                    0 &  0 & -1 & -2 & -2 & -1 & -1 & -2 & -2 \\
            \end{matrix}\right\rrbracket$ 
                      & $\begin{pmatrix}
                          11 \\ 0 \\ 1 \\ 2 \\6
                      \end{pmatrix}$ \\ \\
        $\left[\begin{matrix}
                     1 &  1 & 1 & 0 & 0 & 0 & 0 & 0 & 0 \\
                     1 &  2 &  1 &  0 &  0 &  0 &  0 &  0 &  0 \\
                     1 &  1 &  1 &  0 &  0 &  0 &  0 &  0 &  0 \\
                     0 &  0 &  0 &  0 & -1 & -1 & -1 & -1 & -1 \\
                     0 &  0 &  0 & -1 & -1 & -1 & -1 & -1 & -1 \\
                     0 &  0 &  0 & -1 & -1 &  0 & -1 & -1 &  0 \\
                     0 &  0 &  0 & -1 & -1 & -1 & -1 & -1 &  0 \\
                     0 &  0 &  0 & -1 & -1 & -1 & -1 &  0 &  0 \\
                     0 &  0 &  0 & -1 & -1 &  0 &  0 &  0 &  1 \\
            \end{matrix}\right\rrbracket$ & 
            $\left[\begin{matrix}
                     1 &  0 &  0 &  0 & -1 & -1 &  0 &  0 &  0\\ 
                     0 &  0 & -1 & -1 & -1 & -1 &  0 &  0 &  0\\ 
                     0 & -1 & -1 & -1 & -1 & -1 &  0 &  0 &  0\\ 
                     0 & -1 & -1 &  0 & -1 & -1 &  0 &  0 &  0\\ 
                    -1 & -1 & -1 & -1 & -1 & -1 &  0 &  0 &  0\\ 
                    -1 & -1 & -1 & -1 & -1 &  0 &  0 &  0 &  0\\ 
                     0 &  0 &  0 &  0 &  0 &  0 &  1 &  1 &  1\\ 
                     0 &  0 &  0 &  0 &  0 &  0 &  1 &  2 &  1\\ 
                     0 &  0 &  0 &  0 &  0 &  0 &  1 &  1 &  1\\ 
            \end{matrix}\right\rrbracket$ 
                      & $\begin{pmatrix}
                          6 \\ 3 \\ 5 \\ 1 \\4
                      \end{pmatrix}$ \\
   &  \end{tabular}
    }
    \label{tab:size9}
\end{table}

\section{Conclusion}
\label{sec:conclusion}

The topological characterisation of a chaotic attractor permits to describe its topological properties. 
The result of this methodology depends on the Poincaré section choice \cite{rosalie2015systematic}.
A framework is provided to compare attractor's templates using the \emph{elementary
mixer} describing the chaotic mechanism of a template bounded by a genus--1
torus. 
The algebraic relations between linking matrices of mixers are used to relate 
linking matrix describing the same chaotic mechanism. 
Chaotic mechanisms are equivalent up to torsions and symmetries: the dynamical properties are the same.
As an example, we thus prove that the Burke-Shaw attractor and the Malasoma attractor have the same chaotic mechanism. 

An iterative algorithm is also provided to generate all possible chaotic mechanisms with their elementary mixers for templates of two to eight strips.
The results are presented in Tabs. \ref{tab:size3}, \ref{tab:size4} and \ref{tab:size5}
containing all possible chaotic mechanisms.
In future work, we plan to search for attractors having these theoretical chaotic mechanisms where some, up to the knowledge of the author, has not been observed yet.
This specific representation with left and right elementary mixers will help to iteratively build a linking matrix during the process of topological characterisation.

In addition to allowing comparison of chaotic attractors, these sets of chaotic mechanisms could be used to classify chaotic attractor by complexity. 
When chaotic attractors are used in algorithms for their dynamical properties \cite{rosalie2018chaos, Yildirim2023} or for benchmarking algorithms \cite{Wu2024, Wang2024}, this is generally done with classical chaotic attractors such as Lorenz and Rossler attractors.
Authors could consider having not only chaos for this purpose, but also distinct chaotic mechanisms with a hierarchy.
The Rössler system is a good candidate because when a parameter is varied, the system exhibits distinct chaotic mechanisms of size 2 to 4 \cite{rosalie2016templates_b} with chaotic mechanisms $a$, $c$ and $f$.

A drawback of the topological characterisation method is when a parameter of a differential equation system is varied: all the computations and analysis required to obtain the template have to be performed again due to the properties of chaotic attractors. 
Recent works \cite{rosalie2016templates_b,Serrano2021} propose methodologies for performing the analysis with a range of parameters where the dynamics signature has similarities (the first return map structure is similar). 
The classification presented here could help to list chaotic mechanisms exhibited by a given system when parameters are varied.
For future work, integrating chaotic mechanisms into the \texttt{cate} software \cite{olszewski2018visualizing} could facilitate dissemination.

\section*{Acknowledgments}

The author would like to thanks his co-authors of the article
\cite{olszewski2018visualizing} for \texttt{cate}: an tool to draw templates
from linking matrix (see \url{https://gitlab.inria.fr/cate/cate}).

\printbibliography

\appendix

\section{Implementation}
\label{app:implementation}

The framework presented in this article is implemented in Python. The class Mixer provides the necessary tools to perform algebraic operations on linking matrices (source code available at: \url{https://github.com/mart1rosalie/chaotic_mechanism}).
Python scripts are also provided for the examples (\texttt{example\_1.py} and \texttt{example\_2.py}) as well as for generating chaotic mechanisms:
\begin{itemize}
    \item \texttt{list\_of\_chaotic\_mechanisms.py}
    \item \texttt{extension\_by\_concatenation.py} whose pseudocode is given in Alg.~\ref{algo:extension} using properties of the \texttt{set} Python object.
\end{itemize}

\begin{algorithm}[htp]
\caption{Generation of sets of elementary matrices}
\KwIn{Initial mixer $a$}
\KwOut{List of sets of elementary matrices of increasing sizes}

Initialize $list\_of\_set \gets \{ \{a\} \}$\;
\For{$size \gets 3$ \KwTo $8$}{
    $list\_of\_candidates \gets \emptyset$\;
    \ForEach{$elementary \in list\_of\_set[size-3]$}{
        $temp \gets \text{submixers from } a + elementary \text{ of size }= size$\;
        \ForEach{$submixer \in temp$}{
            $list\_of\_candidates \gets list\_of\_candidates \cup \{submixer\}$\;
        }
        $temp \gets \text{submixers from } a^p + elementary \text{ of size }= size$\;
        \ForEach{$submixer \in temp$}{
            $list\_of\_candidates \gets list\_of\_candidates \cup \{submixer\}$\;
        }
    }
    Append $\emptyset$ to $list\_of\_set$\;
    \ForEach{$candidate \in list\_of\_candidates$}{
        $list\_of\_set[size-2] \gets list\_of\_set[size-2] \cup \{ elementary\_of(candidate) \}$\;
    }
}
\label{algo:extension}
\end{algorithm}

\end{document}